\documentclass[12pt,review,sort&compress]{elsarticle}

\usepackage{amssymb}
\usepackage{amsmath}
\usepackage{graphicx}
\usepackage{tabularx}
\usepackage{cite}
\usepackage{color}
\usepackage{fullpage}
\usepackage{subfigure}
\usepackage{subfigmat}
\usepackage{etoolbox}
\usepackage{url}
\usepackage{float}
\usepackage{bm}
\usepackage[mediumspace,mediumqspace,squaren]{SIunits}
\usepackage[dvipsnames]{xcolor}
\usepackage[title]{appendix}

\newcommand{\ve}[1]{\mathbf{#1}}

\newcommand{\bu}{\ve{u}}
\newcommand{\bn}{\ve{n}}

\newcommand{\bq}{\ve{q}}
\newcommand{\bF}{\ve{F}}
\newcommand{\bL}{\ve{L}}
\newcommand{\br}{\ve{r}}
\newcommand{\bh}{\ve{h}}

\newcommand{\bI}{\ve{I}}

\newcommand{\eps}{\varepsilon}

\newcommand\kdivu{K \nabla \cdot \bu}

\journal{J. Comp. Phys.}

\begin{document}

\begin{frontmatter}

\title{
    An assessment of multicomponent flow models and interface capturing schemes for spherical bubble dynamics
}

\author{Kevin Schmidmayer\corref{cor}}
\cortext[cor]{Corresponding author}
\ead{kevinsch@caltech.edu}
\author{Spencer H. Bryngelson\corref{none}}
\ead{spencer@caltech.edu}
\author{Tim Colonius\corref{none}}
\ead{colonius@caltech.edu}

\address{
    Division of Engineering and Applied Science, California Institute of Technology \\  
    1200 E.\ California Blvd., Pasadena, CA 91125, USA
}

\begin{abstract}
    Numerical simulation of bubble dynamics and cavitation is challenging;
    even the seemingly simple problem of a collapsing spherical bubble is difficult to 
    compute accurately with a general, three-dimensional, compressible, multicomponent flow solver.  
    Difficulties arise due to both the physical model and the numerical method chosen for its solution. 
    We consider the 5-equation model of~\citet{allaire},
    the 5-equation model of~\citet{kapila2001}, and the 
    6-equation model of~\citet{relaxjcp} as candidate approaches for 
    spherical bubble dynamics, and both MUSCL and WENO interface-capturing methods 
    are implemented and compared.
    We demonstrate the inadequacy of the traditional 
    5-equation model of~\citet{allaire} for spherical bubble collapse problems and explain 
    the corresponding advantages of the augmented model of~\citet{kapila2001}
    for representing this phenomenon. Quantitative comparisons between the augmented
    5-equation and 6-equation models for three-dimensional bubble collapse problems 
    demonstrate the versatility of pressure-disequilibrium models.
    Lastly, the performance of pressure disequilibrium model 
    for representing a three-dimensional spherical bubble collapse for different bubble 
    interior/exterior pressure ratios is evaluated for different numerical methods.
    Pathologies associated with each factor and their 
    origins are identified and discussed.
\end{abstract}


\end{frontmatter}

\section{Introduction}

Amongst other features, cavitation involves the growth and collapse of a gas bubble in a liquid.
Many applications require a detailed understanding of this process 
in or near soft materials, including biological tissues for
medical purposes~\citep{
brennen2015medicine,
marmottant2019softAndBiologicalMatter,
estrada2018high, adams2018biomaterial,
ando2018cavitationInGel} and  
polymeric coatings and biofouling in industry~\citep{leighton2017elasticPlastic}. 
Preliminary studies have shown that bubble dynamics are sensitive to the properties of these 
materials~\citep{barajas2017effects,johnsen2015viscoelastic},
motivating a comprehensive multi-scale theory capable of predicting
complex bubble cavitation.

Before considering the viscoelasticity of soft materials,  
accurate algorithms for bubble dynamics in Newtonian liquids must be developed. 
Indeed, even the seemingly simple problem of a collapsing spherical bubble is challenging to 
compute accurately with general, three-dimensional (3D), fully-compressible computational methods 
for a significant range of bubble/ambient pressure ratios (and thus
interface Mach numbers). Here, we use this problem as a case study for the ability 
of a physical model, and its coupled numerical method, to predict bubble dynamics generally.

Diffuse interface (interface-capturing) methods appear to be well-suited for this problem 
when compared to other interface tracking and 
capturing schemes~\citep{petitpas2008phdthesis, fuster2018review}.
Such methods combine a multicomponent flow model with shock-capturing finite volume
methods. Their discrete-level conservation allows 
the compressibility of all phases and mixtures
to be represented on the computational grid and interfaces 
appear and vanish naturally, irrespective of their corresponding density ratio.
Herein, we assess the difficulties that arise during a spherical bubble collapse 
from the physical multicomponent flow model and its coupling to the numerical method.

The mechanical-equilibrium multicomponent model of~\citet{allaire} has been widely
used and can faithfully represent shock-induced collapses~\citep{coralic2014WENO5,
beig2015maintaining, veilleux2018cav, wang2018ellipsoidalBubble} and droplet 
atomization~\citep{meng2014, meng2018}.
Unfortunately, this model cannot predict the collapse time and minimum radius of
the Rayleigh collapse problem~\citep{tiwari2013diffuse,
koumoutsakos2017large}. This problem can be averted via the thermodynamically consistent model 
of~\citet{kapila2001}~\citep{tiwari2013diffuse, koumoutsakos2017large},
which includes a term ($\kdivu$) in the volume-fraction evolution equation
to represent compressibility in mixture regions. 
Unfortunately, this additional term leads to numerical instabilities 
during strong compression and expansion near the interface~\citep{relaxjcp, beig2018temperatures}. 
Instead, we propose using a pressure-disequilibrium model~\citep{relaxjcp}, 
which relaxes the phase-specific pressures algorithmically at each time step,
and averts the stability issues of the $\kdivu$ term. 
This model theoretically converges to the 
mechanical-equilibrium model of~\citet{kapila2001} under mesh refinement,
and while it has been utilized
for cavitating flows~\citep{relaxjcp, torpille},
detonating flows~\citep{deto},
surface-tension driven flows~\citep{schmidmayer2017capillary}, 
droplet atomization~\citep{schmidmayer2017phd, schmidmayer2018AMR}, 
and fracture and fragmentation in
ductile materials~\citep{favrie2009solid, ndanou2015fragmentation},
it has not been applied to bubble dynamics
or particularly to collapsing bubbles.

The multicomponent flow models are solved using 
shock-capturing finite-volume schemes and Riemann
solvers for fluxes~\citep{godunovRussian, toro97}.
High-order spatial reconstructions, such as MUSCL~\citep{van1979towards, toro97, schmidmayer2019ecogen}
and WENO~\citep{shu1996efficientWENO, shukla2010interface, coralic2014WENO5, tiwari2013diffuse},
are often used, along with their variants WENO-Z~\citep{zhang2006weighted},
WENO-CU6~\citep{hu2010adaptive, hu2011scale}, and
TENO~\citep{fu2016family}. Herein, we will consider
MUSCL and the WENO of~\citet{shu1996efficientWENO},
coupled with the HLLC approximate Riemann solver~\citep{toro97, coralic2014WENO5, relaxjcp}
as standard approaches for solving the multicomponent flow equations. 
Following usual procedure, these are coupled to total-variation-diminishing
time integrators as an attempt to suppress spurious oscillations at material 
interfaces under refinement~\citep{godunovRussian, van1977MC, toro97, gottlieb1998RK3}.

We first present the diffuse-interface multicomponent models in section~\ref{sec:Models}. 
The numerical methods we employ to solve the resulting equations are outlined 
in section~\ref{sec:Methods}.
The setup of the spherical-bubble-collapse problems we consider 
are presented in section~\ref{sec:setup}.
In section~\ref{sec:effect_noKdivU} we demonstrate and explain the utility 
of the $\kdivu$ term in the mechanical-equilibrium models. The convergence and behavior
of this improved equilibrium model and the usual pressure-disequilibrium model are studied
in section~\ref{sec:comparison_models} for the collapse and rebound of
spherical bubbles. Artifacts of the numerical methods we consider are 
examined in section~\ref{sec:comparison_methods}, including 
an investigation of interface sharpening techniques in section~\ref{s:THINC}.
Finally, the pathologies
identified are discussed in section~\ref{sec:Conclusion}.

\section{Multicomponent flow models}\label{sec:Models}

The compressible multicomponent flow models we present 
can all be written as
\begin{gather}
    \frac{\partial \bq}{\partial t} + \nabla \cdot \bF 
    \left( \bq \right) + \bh \left( \bq \right) 
    \nabla \cdot \bu = 
    \br \left( \bq \right),
    \label{eq_general}
\end{gather}
where $\bq$ is the state vector, $\bF$ is the flux tensor, 
$\bu$ is the velocity field, and $\bh$ and $\br$ are 
non-conservative quantities we describe subsequently.
We only consider mechanical-equilibrium models that formally
conserve mass, momentum, and total energy, and neglect the effects of 
viscosity, phase change and surface tension.

\subsection{Mechanical-equilibrium model of Allaire et al.~\citep{allaire}}

We first consider the mechanical-equilibrium model of~\citet{allaire},
which we call the 5-equation model. For a two-phase flow, we have
\begin{gather}
    \bq = \left[ \begin{array}{c}
        \alpha_1 \\
        \alpha_1 \rho_1 \\
        \alpha_2 \rho_2 \\
        \rho \bu \\
        \rho E
    \end{array} \right], \quad
    \bF = \left[ \begin{array}{c}
        \alpha_1 \bu \\
        \alpha_1 \rho_1 \bu \\
        \alpha_2 \rho_2 \bu \\
        \rho \bu \otimes \bu + p \bI \\
        \left( \rho E + p \right) \bu
    \end{array} \right], \quad 
    \bh = \left[ \begin{array}{c}
        - \alpha_1 \\
        0 \\
        0 \\
        \mathbf{0} \\
        0
    \end{array} \right], \quad
    \br = \left[ \begin{array}{c}
        0 \\
        0 \\
        0 \\
        \mathbf{0} \\
        0
    \end{array} \right],
    \label{model_allaire}
\end{gather}
where $\rho$, $\bu$, and $p$ are the mixture 
density, velocity, and pressure, respectively,
and $\alpha_k$ is the volume fraction, for which $k$ indicates the phase index. 
The mixture total energy is 
\begin{gather}
    E = e + \frac{1}{2} \| \bu \|^2,
\end{gather} 
where $e$ is the mixture specific internal energy
\begin{gather}
    e = \sum_{k=1}^2 Y_k e_k \left( \rho_k , p \right).
    \label{eq_internalEnergy}
\end{gather}
In~\eqref{eq_internalEnergy}, $e_k$ is defined
via an equation of state and $Y_k$ are the mass fractions
\begin{gather}
    Y_k = \frac{\alpha_k \rho_k}{\rho} .
\end{gather}
Herein, we will consider a two-phase mixture of gas ($g$) and liquid ($l$),
for which the gas is modeled by the ideal-gas equation of state
\begin{gather}
    p_g = ( \gamma_g - 1) \rho_g e_g ,
\end{gather}
where $\gamma_g = 1.4$, and the liquid is modeled by the stiffened-gas equation of state
\begin{gather}
    p_l = ( \gamma_l - 1) \rho_l e_l - 
    \gamma_l \pi_\infty,
\end{gather}
where $\gamma_l$ and $\pi_\infty$ are case-specific model parameters~\citep{SGEOS}.
The mixture quantities are 
\begin{gather}
    \rho = \sum_{k=1}^2 \alpha _k \rho_k  \quad \text{and} \quad
    p = \sum_{k=1}^2 \alpha _k p_k,
\end{gather}
and 
\begin{gather}
    \rho c^2 = \sum_{k=1}^2 \frac{\alpha_k \rho_k c_k^2}{\beta \left( \gamma_k - 1 \right)} , 
    \quad 
    \beta = \sum_{k=1}^2 \frac{\alpha_k}{\gamma_k - 1} ,
    \label{eq_speedOfSound_Allaire}
\end{gather}
where $c$ is the mixture speed of sound, and $c_k$ and $\gamma_k$ are the 
speed of sound and polytropic coefficient of phase $k$.
We note that while this model conserves mass, momentum, and total energy,
it does not strictly obey the second law of thermodynamics~\citep{allaire, schmidmayer2017capillary}.

\subsection{Mechanical-equilibrium model of~\citet{kapila2001}}

The thermodynamically consistent
mechanical-equilibrium model of~\citet{kapila2001},
which we call the 5-equation model with $\kdivu$, has
\begin{gather}
    \bq = \left[ \begin{array}{c}
        \alpha_1 \\
        \alpha_1 \rho_1 \\
        \alpha_2 \rho_2 \\
        \rho \bu \\
        \rho E
    \end{array} \right], \quad
    \bF = \left[ \begin{array}{c}
        \alpha_1 \bu \\
        \alpha_1 \rho_1 \bu \\
        \alpha_2 \rho_2 \bu \\
        \rho \bu \otimes \bu + p \bI \\
        \left( \rho E + p \right) \bu
    \end{array} \right], \quad 
    \bh = \left[ \begin{array}{c}
        - \alpha_1 \\
        0 \\
        0 \\
        \mathbf{0} \\
        0
    \end{array} \right], \quad
    \br = \left[ \begin{array}{c}
        \kdivu \\
        0 \\
        0 \\
        \mathbf{0} \\
        0
    \end{array} \right],
    \label{model_kapila}
\end{gather}
where only $\br$ is different from~\eqref{model_allaire}. Here, 
$K$ is
\begin{gather}
    K = \frac{\rho _2 c_2^2 - \rho _1 c_1^2}{\frac{\rho _2 c_2^2}{\alpha _2} + \frac{\rho _1 c_1^2}{\alpha _1}},
\end{gather}
and $\kdivu$ represents 
expansion and compression of each phase in 
mixture regions.
In this case, the mixture speed of sound follows from 
\begin{gather}
    \frac{1}{\rho c^2} = \sum_{k=1}^2 \frac{\alpha_k}{\rho_k c_k^2},
    \label{eq_speedOfSound_Wood}
\end{gather}
which is also the Wood speed of sound~\citep{wood, wallis1969}. 

\subsection{Pressure-disequilibrium model of~\citet{relaxjcp}}

The pressure-disequilibrium model of~\citet{relaxjcp}, 
which we call the 6-equation model, is expressed as
\begin{gather}
    \bq = \left[ \begin{array}{c}
        \alpha_1 \\
        \alpha_1 \rho_1 \\
        \alpha_2 \rho_2 \\
        \rho \bu \\
        \alpha _1 \rho _1 e_1 \\
        \alpha _2 \rho _2 e_2
    \end{array} \right], \quad
    \bF = \left[ \begin{array}{c}
        \alpha_1 \bu \\
        \alpha_1 \rho_1 \bu \\
        \alpha_2 \rho_2 \bu \\
        \rho \bu \otimes \bu + p \bI \\
        \alpha _1 \rho _1 e_1 \bu \\
        \alpha _2 \rho _2 e_2 \bu
    \end{array} \right], \quad 
    \bh = \left[ \begin{array}{c}
        - \alpha_1 \\
        0 \\
        0 \\
        \mathbf{0} \\
        \alpha _1 p_1 \\
        \alpha _2 p_2
    \end{array} \right], \quad
    \br = \left[ \begin{array}{c}
        \mu \delta p \\
        0 \\
        0 \\
        \mathbf{0} \\
        - \mu p_I \delta p \\
        \mu p_I \delta p
    \end{array} \right],
    \label{model_relax}
\end{gather}
where $\br$ represents the relaxation of pressures between the phases with 
coefficient $\mu$. The interfacial pressure is 
\begin{gather}
    p_I =\frac{z_2 p_1 + z_1 p_2}{z_1 + z_2},
\end{gather}
where ${z_k} = {\rho _k}{c_k}$ is the acoustic impedance of the phase $k$,
and
\begin{gather}
    \delta p = p_1 - p_2,
\end{gather}
is the pressure difference between the two phases. Since $p_1 \neq p_2$ here, 
the total energy equation of the mixture is replaced by the internal-energy equation for each phase.
Nevertheless, conservation of the mixture total energy can be written in its usual form
\begin{gather}
    \frac{\partial \rho E}{\partial t} + \nabla \cdot \left[ \left( \rho E + p \right) \bu \right] = 0 .
    \label{eq_total_energy}
\end{gather}
We note that~\eqref{eq_total_energy} is redundant when the internal energy equations are also 
computed. However, in practice we include it in our computations to ensure that 
the total energy is numerically conserved, and thus preserve a 
correct treatment of shock waves (more details can be found in~\citet{relaxjcp}).

The mixture speed of sound is defined according to
\begin{gather}
    c^2 = \sum_{k=1}^2 Y_k c_k^2 .
    \label{eq_speedOfSound_Saurel}
\end{gather}
After applying the infinite pressure-relaxation procedure detailed in
section~\ref{sec_relaxationProcedure}, 
the effective mixture speed of sound matches~\eqref{eq_speedOfSound_Wood}.
We will discuss the influence of sound speed for 
interface problems in section~\ref{sec:results_OutOfEq}.

\section{Numerical methods}\label{sec:Methods}

We solve~\eqref{eq_general} numerically; the time evolution of $\bq$
on a computational cell $i$ with volume $V_i$ and surface $A$ with normal unit vector 
$\mathbf{n}$ is given by the explicit finite-volume~\citet{godunovRussian} scheme
\begin{gather}
    \bq^{n+1}_i = \bq^{n}_i - 
    \frac{\Delta t}{V_i} \left( \sum^N_{s=1} A_s  \bF_s^\star \cdot \bn_s + 
    \bh \left( \bq_i^n \right) \sum^N_{s=1} A_s  \bu^{\star}_{s} \cdot \mathbf{n}_s \right) ,
    \label{scheme_firstOrder}
\end{gather}
where $n$ is the time-step index. Here, we label this basic first-order-accurate finite-volume scheme as 
\texttt{FV1}. At the volume--volume interfaces, the associated Riemann
problem is computed using the HLLC approximate solver~\citep{coralic2014WENO5, relaxjcp, toro97},
giving the flux tensor and flow-velocity vector
$\bF^\star_s$ and $\bu^\star_s$, respectively.
The solution of~\eqref{scheme_firstOrder} is restricted by the usual CFL criterion.

\subsection{Spatial and time reconstruction}

Herein, we utilize both MUSCL and WENO spatial reconstructions.
We use the second-order-accurate MUSCL scheme of \citet{schmidmayer2019ecogen} 
(labeled here as \texttt{MUSCL2}) with two-step time  integration
\begin{align}
        {\bq^{n+\frac{1}{2}}_i} & {= \bq^{n}_i + \frac{1}{2} \Delta t \mathbf{L} \left( \bq_i^n \right) ,} \\ 
        {\bq^{n+1}_i} & {= \bq^{n}_i + \Delta t \mathbf{L} \left( \bq_i^{n+\frac{1}{2}} \right) ,}
        \label{scheme_muscl}
\end{align}
where $\bL$ is the numerically approximated fluxes
and non-conservative terms. The first step is a prediction for the second step and the usual 
piece-wise linear MUSCL reconstruction~\citep{toro97} is used on the primitive variables.
The monotonized central (MC)~\citep{van1977MC} slope limiter is employed as an attempt to minimize interface diffusion
and its behavior is investigated in section~\ref{sec:comparison_methods}.
This method has been previously implemented for the pressure-disequilibrium 
model~\citep{relaxjcp, torpille, deto, schmidmayer2017capillary, 
schmidmayer2019ecogen, schmidmayer2018AMR, schmidmayer2018ecogen3AF, 
ndanou2015fragmentation}.

The WENO scheme is either third- or fifth-order accurate (labeled here as \texttt{WENO3} and \texttt{WENO5}, respectively) 
and reconstructs the primitive variables~\citep{coralic2014WENO5}.
In this case, the time derivative is computed via the third-order 
TVD Runge--Kutta algorithm~\citep{gottlieb1998RK3}
\begin{align}
        {\bq^{(1)}_i} & {= \bq^{n}_i + \Delta t \mathbf{L} \left( \bq_i^n \right) ,} \label{scheme_weno1}\\
        {\bq^{(2)}_i} & {= \frac{3}{4} \bq^{n}_i + \frac{1}{4} \bq^{(1)}_i + \frac{1}{4} \Delta t \mathbf{L} \left( \bq_i^{(1)} \right) ,} \\
        {\bq^{n+1}_i} & {= \frac{1}{3} \bq^{n}_i + \frac{2}{3} \bq^{(2)}_i + \frac{2}{3} \Delta t \mathbf{L} \left( \bq_i^{(2)} \right) .} \label{scheme_weno3}
\end{align}
This method has previously been implemented for the pressure-equilibrium models 
of~\citet{allaire} 
\citep{coralic2014WENO5, beig2015maintaining, meng2018, meng2014, veilleux2018cav} 
and of~\citet{kapila2001} 
\citep{tiwari2013diffuse, koumoutsakos2017large, rodriguez2018viscoelasticity};
here, we also utilize it for the pressure-disequilibrium model of~\citet{relaxjcp}. 

\subsection{Pressure-relaxation procedure}\label{sec_relaxationProcedure}

The pressure-disequilibrium model~\eqref{model_relax} requires pressure-relaxation
to converge to a single, equilibrium pressure. We use the infinite-relaxation
procedure of~\citet{relaxjcp}. At each time step it solves the non-relaxed, 
hyperbolic equations ($\mu \to 0$) using~\eqref{scheme_firstOrder},
then relaxes the disequilibrium pressures
for $\mu \to +\infty$. The latter is combined with a re-initialization procedure 
to ensure the conservation of total energy, and thus converges to the 
mechanical-equilibrium model of~\citet{kapila2001}~\eqref{model_kapila}.
When multi-stage time integration is used, the relaxation procedure is performed at each 
stage. Thus, there is only one pressure at the end of each stage and 
the reconstructed variables are the same for all models.  As a result,
simulations of the pressure-disequilibrium model are only about 
$5\%$ more expensive than the models of~\citet{allaire} and~\citet{kapila2001} for
the spherical-bubble-collapse cases we consider subsequently.

\section{Setup of the spherical-bubble-collapse problem}\label{sec:setup}

\begin{figure}
    \centering
    \includegraphics{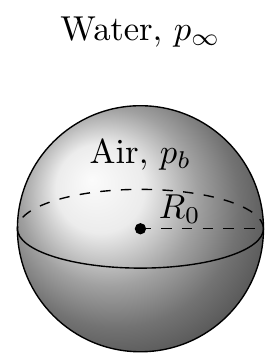}
    \caption{Problem configuration for a collapsing spherical bubble.}
    \label{f:sphere_setup}
\end{figure}

As a step towards understanding the practical
differences between the presented models and methods,
we consider the behavior of a collapsing spherical bubble. 
The problem setup is shown in Figure~\ref{f:sphere_setup}.
We initialize the bubble with radius $R_0$ and
the computational domain has size $L = 320 R_0$, which is sufficiently large 
to avoid boundary effects.
Initially, the bubble has a uniform internal pressure $p_b$, and the exterior pressure
increases gradually up to the far-field pressure $p_\infty$ 
according to the Rayleigh--Plesset 
equation~\citep{brennen1995cavitation, tiwari2013diffuse}:
\begin{gather}
    p(R) = 
        p_\infty + \frac{R_0}{R} \left( p_b - p_\infty \right) .
\end{gather}
In the following, this pressure initialization is labeled as initial interface equilibrium with $\dot R_0 = 0$.
We consider cases with both modest and high initial pressure ratios,
as shown in Table~\ref{t:cases}.
The water is parameterized by $\gamma_l = 2.35$ and 
$\pi_\infty = \unit{10^9}{\pascal}$~\citep{SGEOS, olivier2005, evap, torpille, pelanti2014mixture, beig2015maintaining}.
\begin{table}[H]
    \centering
    \begin{tabular}{|l|c|c|c|}
        \hline
        \multicolumn{1}{|c|}{Case}    & $p_\infty$ [Pa] & $p_b$ [Pa] & $p_\infty/p_b$ \\
        \hline
        \textbf{1}: Low-pressure-ratio      & $10^5$          & $10^4$ & 10      \\
        \hline
        \textbf{2}: High-pressure-ratio      & $5 \times 10^6$ & $3550$ & 1427 \\
        \hline
    \end{tabular}
    \caption{Nominal initial conditions for the cases simulated.}
    \label{t:cases}
\end{table}

We simulate the flow on a cubical, rectilinear grid with
$N_{R_0}$ nodes in each coordinate direction per initial bubble radius near the bubble ($R \leqslant 1.5 R_0$);
far from the bubble  ($R > 1.5 R_0$), the grid is stretched nonuniformly to 
accommodate the large computational domain $L$.
To reduce computational cost, one octant of the domain is
computed, with symmetry boundary conditions mimicking the bubble dynamics
in neighboring regions. We performed two simulations for each pressure
ratio, one without mesh stretching and another with the complete physical
domain (no symmetry boundary conditions), and compared them against the
simulations presented hereafter
to confirm that our results are insensitive to
both of these procedures.

When using the \texttt{WENO5} method, the bubble interface is 
smeared in the radial direction over few grid cells.
The smearing procedure is commonly employed in multicomponent
models when fifth-order WENO reconstruction is used~\citep{johnsen2006WENO,
johnsen2009numerical,
niederhaus2008computational,
shukla2010interface,
tiwari2013diffuse,
coralic2014WENO5,
tiwari2015growth,
beig2015maintaining,
koumoutsakos2017large,
wang2017numerical,
veilleux2018cav, beig2018temperatures},
as it appears that unphysical oscillations or numerical instabilities can
occur without it.
The initial interface smearing procedure we employ involves smearing the volume fraction
across the interface using
an hyperbolic tangent function~\citep{tiwari2013diffuse}
\begin{gather}
     \alpha_g = \frac{1}{2} \left[ 1 - \textrm{tanh} \left( \frac{R - R_0}{2 D} \right) \right] ,
\end{gather}
where $D$ is the characteristic length of the corresponding computational cell;
the conservative variables then follow from simple mixture relations,
allowing thermodynamic consistency.
The physical artifacts associated with this procedure are discussed
in section~\ref{sec:results_OutOfEq}.

In the following, we use the radial bubble-wall evolution to compare the performance of the three different multicomponent models. 
We define an effective bubble radius, $R$, as
\begin{gather}
    R = \left( \frac{3 V_b}{4 \pi} \right)^{\frac{1}{3}}, \quad \text{where} \quad
    V_b = \sum_{i=1}^N \alpha_{g,i} V_{c,i}
\end{gather}
is the total volume of the gas phase, $N$ is the total number of grid cells, 
and $\alpha_{g,i}$ and $V_{c,i}$ are the gas volume fraction and the volume of cell $i$, respectively. 
The radial bubble-wall evolution is presented in a non-dimensionalized form where
\begin{gather}
    t_c = 0.915 R_0 \sqrt{ \frac{\rho_l}{p_\infty} }
\end{gather}
is the nominal total collapse time from its initial (maximum) radius $R_0$~\citep{brennen1995cavitation}.
In our implementation, we compute about $69 \times 10^3$ and $18 \times 10^3$ time steps per $t_c$ for
cases \textbf{1} and \textbf{2}, respectively. 


\section{Results}\label{sec:results}

\subsection{Effect of $\kdivu$ on the 5-equation model}\label{sec:effect_noKdivU}

We first reconsider the behavior and influence of the $\kdivu$ term from the 5-equation models
on the spherical-bubble-collapse problem using the \texttt{WENO5} scheme as
previously presented by \citet{tiwari2013diffuse}.

\begin{figure}
    \centering
    \includegraphics[width=0.995\columnwidth]{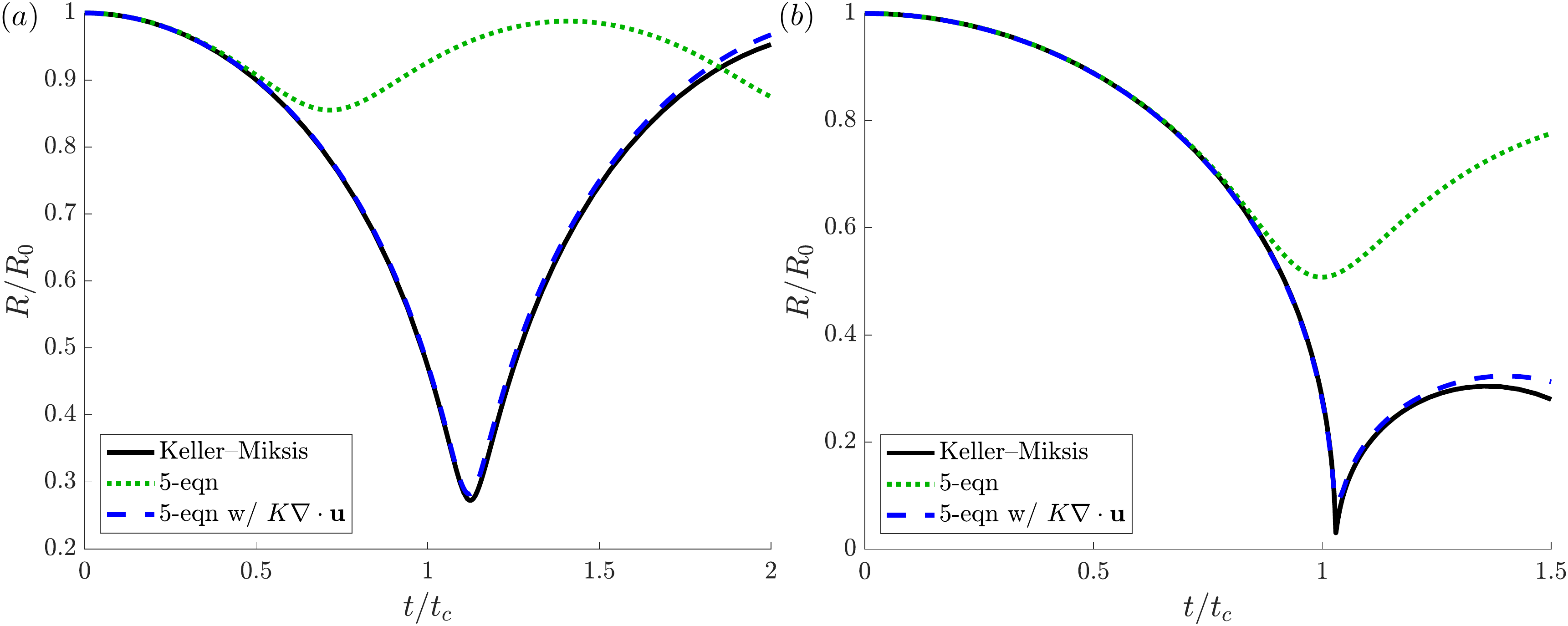}
    \caption{
    Radial bubble-wall evolution for 
    (a) $p_\infty / p_b = 10$ with $N_{R_0} = 25$ and
    (b) $p_\infty / p_b = 1427$ with $N_{R_0} = 50$.
    Solutions are computed using the 5-equation models with \texttt{WENO5}
    as well as the Keller--Miksis equation.
    }
    \label{fig_5eq_noKdivU}
\end{figure}

Figure \ref{fig_5eq_noKdivU} shows that in both pressure-ratio cases,
only the model with $\kdivu$
agrees with a nominal exact solution following the 
Keller--Miksis equation~\citep{keller1980bubble}; a compressible form of 
the Rayleigh--Plesset equation. In this case,
the initial interface smearing does not affect this agreement.

The inability of the 5-equation model without $\kdivu$ to 
represent spherical bubble collapse was previously observed
by~\citet{tiwari2013diffuse}, who attributed the better results to enforcement of the second law of thermodynamics~\citep{kapila2001}. We seek here an alternative explanation in terms of the dynamics.

\begin{figure}
    \centering
    \includegraphics[width=0.995\columnwidth]{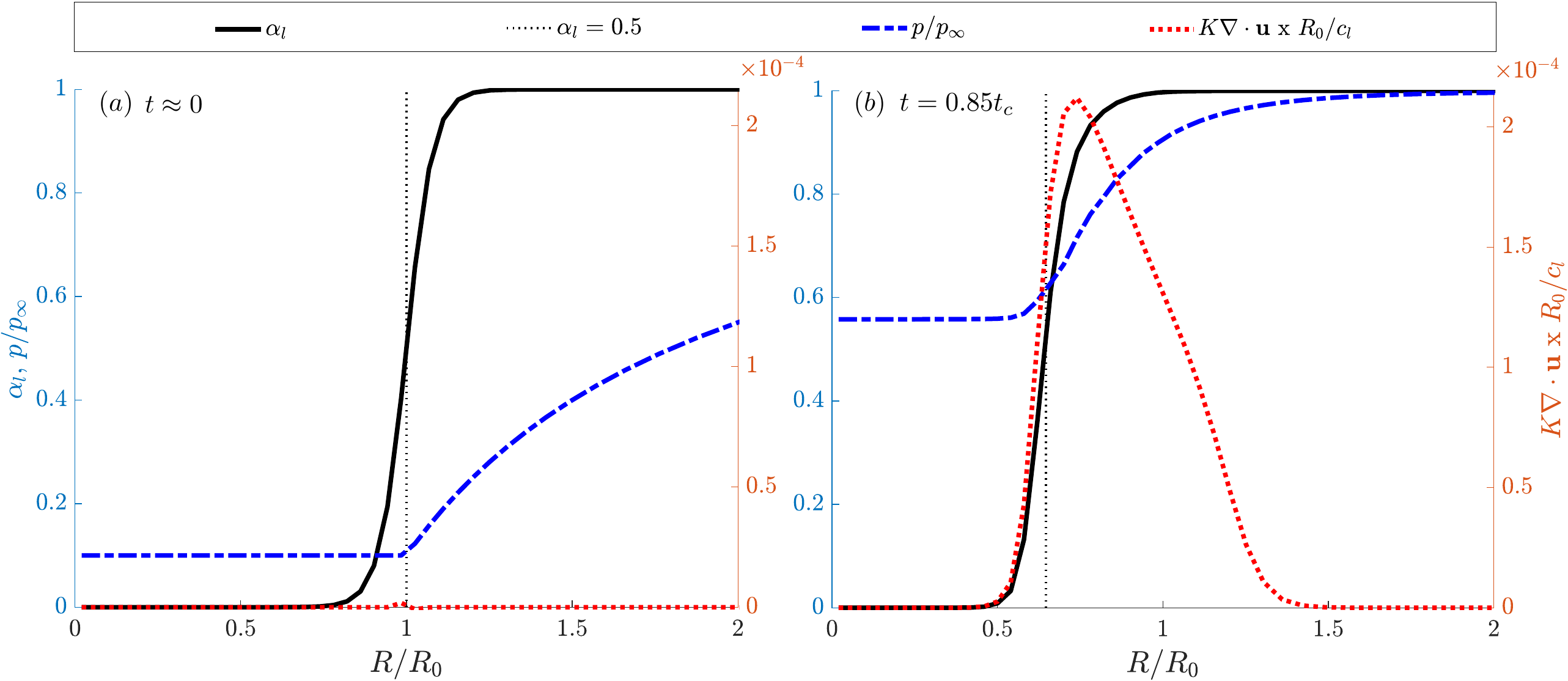}
    \caption{
    Liquid volume-fraction, pressure, and $\kdivu$ for varying radial position $R$
    and the case $p_\infty / p_b = 10$ and $N_{R_0} = 25$ using the 5-equation model
    with $\kdivu$.  Times (a) $t \approx 0$ and (b) $t = 0.85 t_c$ are shown.
    }
    \label{fig_KdivU}
\end{figure}

Figure \ref{fig_KdivU} shows key quantities and 
$\kdivu$ along a radial coordinate at two instances in time.
For $t \approx 0$, the initial interface smearing results in a mixture
region at the interface, for which $K \ne 0$, but $\kdivu \approx 0$ 
because $\nabla \cdot \bu \approx 0$. However, during the
collapse, $\nabla \cdot \bu \ne 0$ and the fluid volume
fractions are modified. We see that $\kdivu$ is positive and is larger 
on the liquid side of the interface. As a result, the liquid volume fraction
increases faster, particularly on the liquid side of the interface, 
than it would without the $\kdivu$ term.
This keeps the interface relatively sharp and results in the larger interface velocity
observed in Figure~\ref{fig_5eq_noKdivU}.

This behavior can be explained via the bubble pressure evolution. 
Initially, the pressure is small inside the bubble and increases
gradually outside of it. During the collapse, the bubble pressure increases,
which reduces the bubble volume due to compression of the gas.
This is dynamically coupled to the interface and intensifies the
collapse. Further, the pressure always increases in the radial direction.
Thus, the gas in the mixture region is more highly compressed 
on the water side of the bubble interface, and so its volume fraction 
decreases more rapidly. This process also intensifies the bubble collapse.
Note that this second effect is not present into the 5-equation model
without $\kdivu$, since this term accounts for expansion and compression
in mixture regions.

\subsection{Comparison of the 5- and 6-equation models}\label{sec:comparison_models}

While the 5-equation model with $\kdivu$ can accurately represent spherical bubble 
dynamics in some cases, it is also often numerically unstable. This is a result of 
significant compression and expansion near the interface, which can occur during 
strong shock or expansion waves. Here, we consider the 
6-equation model as a potential solution to this issue;
under infinite pressure-relaxation, it  
theoretically converges to the 5-equation model with $\kdivu$.
However, when discretized, the equation sets are different and equivalence has neither been demonstrated for high-order schemes, such 
as the \texttt{WENO5} method we consider, nor for the challenging spherical-bubble-collapse test problems.
To test our implementation
and confirm their convergence to one another,
comparison between these methods are presented
for shock tube and vacuum problems in Appendix~\ref{a:shocktube} and \ref{a:vacuum}, and 
consider the collapse of a spherical air bubble in water next.

\begin{figure}
    \centering
    \includegraphics[width=0.995\columnwidth]{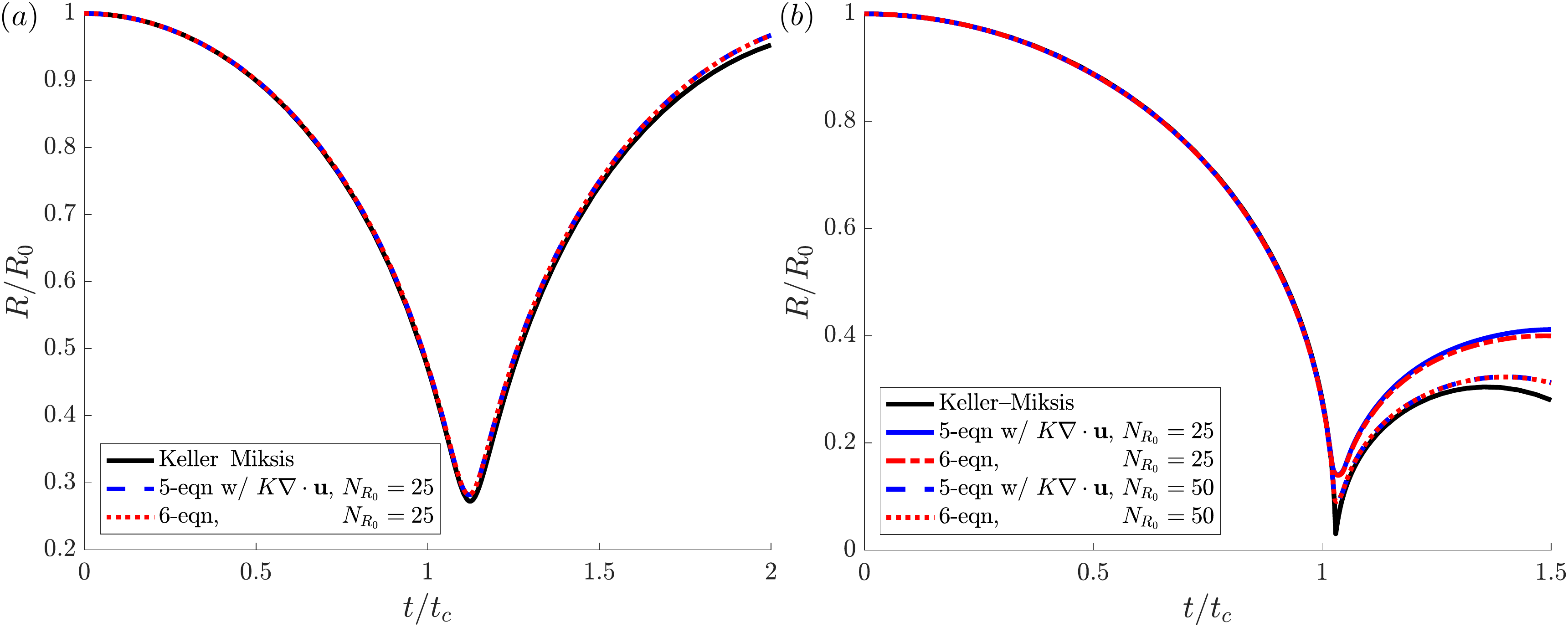}
    \caption{
    Radial bubble-wall evolution for 
    (a) $p_\infty / p_b = 10$ and
    (b) $p_\infty / p_b = 1427$.
    Solutions are computed using \texttt{WENO5}.
    }
    \label{f:sphcollapse}
\end{figure}

Simulation results for nominal low and high pressure ratios are shown in 
Figure~\ref{f:sphcollapse} for both the 5-equation with $\kdivu$ 
and 6-equation models, following the previous subsections.
Both methods agree closely for both cases with the analytic solution of the 
Keller--Miksis equations~\citep{keller1980bubble}, which are 
initialized at equilibrium with $\dot{R}_0 = 0$.
The high pressure ratio case of Figure~\ref{f:sphcollapse} (b)
also shows the spatial convergence of the models.

Since the 6-equation model has closely matched 
the 5-equation model with $\kdivu$ 
for all three of our challenging test cases,
we consider it a potential surrogate to the 5-equation model
that does not inherit its stability issue.
Next, we investigate the behavior of the 6-equation model
when solved by numerical schemes of different character and accuracy.

\subsection{Numerical schemes for the 6-equation model}\label{sec:comparison_methods}

The 6-equation model can be solved via a number of different interface-capturing
numerical methods. We compute its solution using the methods described
in section~\ref{sec:Methods} for a collapsing spherical bubble of varying
initial pressure ratio and interface states as a critical assessment of the viability
of the numerical schemes for cavitating flows.

\subsubsection{Spherical bubble collapse with initial interface equilibrium}\label{sec:results_methods_interfaceEquilibrium}


\begin{figure}
    \centering
    \includegraphics[width=0.995\columnwidth]{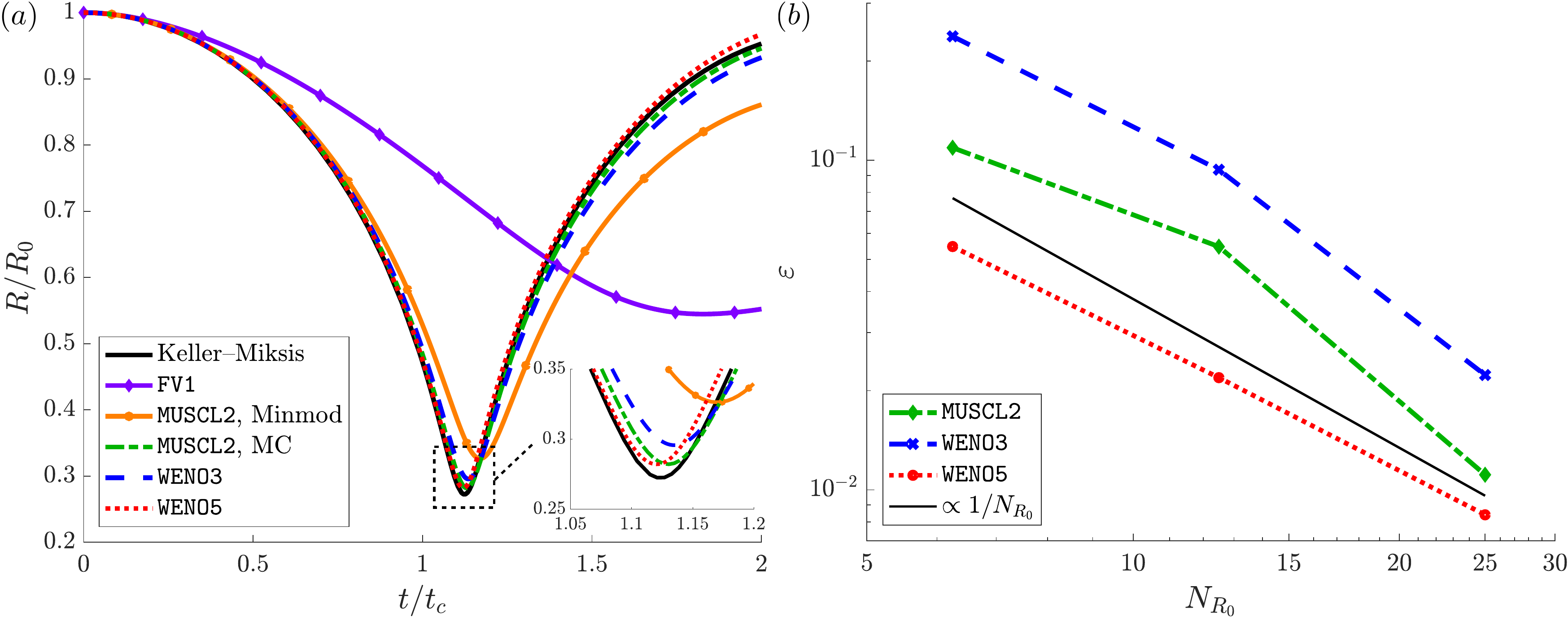}
    \caption{
    Radial bubble-wall evolution for initial interface equilibrium and $p_\infty / p_b = 10$.
    (a) Results for all schemes and flux limiters as labeled (fixed resolution $N_{R_0} = 25$) and
    (b) spatial convergence of the numerical methods.
    }
    \label{fig_allOrders}
\end{figure}

We first consider the case of initial interface equilibrium, $\dot{R}_0 = 0$.
Figure~\ref{fig_allOrders} (a) shows the interface evolution for 
$p_\infty / p_b = 10$, spatial resolution $N_{R_0} = 25$ and for 
the \texttt{FV1}, \texttt{MUSCL2}, \texttt{WENO3}
and \texttt{WENO5} schemes. In addition to the 
MC~\citep{gammie2003MC} slope limiter, the Minmod~\citep{sweby1984superbee, toro97} limiter
is implemented for the \texttt{MUSCL2} scheme (note
that the MC limiter is used when not specified for the \texttt{MUSCL2} scheme).
Here, the slope limiters attempt to reduce numerical dissipation of the scheme. 
As mentioned in section~\ref{sec:setup}, the  
interface is initially smeared for the \texttt{WENO5} cases to guarantee numerical stability.
However, the \texttt{MUSCL2} and \texttt{WENO3} schemes do not require this procedure to remain stable, 
and thus all interfaces are kept sharp at the grid level in these cases.
In Figure~\ref{fig_allOrders} (a) we see that the MC slope limiter performs 
significantly better than the Minmod limiter and similarly for the \texttt{WENO3} scheme,
although the corresponding results are still less accurate than those of 
the \texttt{WENO5} scheme.

To confirm that numerical dissipation is the cause of the discrepancy
between the results of the \texttt{MUSCL2}, \texttt{WENO3} and \texttt{WENO5} schemes, 
we consider the spatial convergence of the numerical methods.
In Figure~\ref{fig_allOrders} (b), this convergence is presented in terms of 
the discrete $L_2$ error $\eps$ as
\begin{gather}
    \eps = \frac{1}{N_t} \sum_{i=0}^{N_t} \frac{ \lVert R(t_i) - R_{KM}(t_i) \rVert}{R_{KM}(t_i)} ,
    \label{e:error}
\end{gather}
where $N_t$ is the number of time steps in the temporal window $t \in [0,2 t_c]$, 
and $R(t_i)$ and $R_{KM}(t_i)$ are the bubble radius at time $t_i$ of our simulations and the 
Keller-Miksis solution, respectively. We see that all methods converge at first order,
matching the expected rate for the numerical solution of flows with
discontinuities~\citep{godunovRussian, vanLeer1974, toro97}.
The \texttt{WENO5} method has the smallest $\eps$, and so we conclude that for small initial pressure ratios
higher-order reconstructions have smaller errors as they suppress numerical diffusion.
In this case, the interface smearing procedure we employ for the \texttt{WENO5} scheme has no apparent consequence on simulation accuracy.


For the flow configurations we consider, the spherical bubble interface is known to be 
physically stable~\citep{frost1986effects, brennen1995cavitation}, and so
non-spherical interfaces are an artifact of the numerical method; 
we use this property to assess the performance of the numerical methods.
The bubble sphericity is computed as~\citep{wadell1935sphericity}
\begin{gather}
    \Psi = \frac{\pi^{\frac{1}{3}} \left( 6 V_b^\prime \right)^{\frac{2}{3}}}{A_b},
\end{gather}
which is the ratio of the surface area of a sphere 
with the same volume as the bubble $V_b^\prime$,
to the surface area of the bubble $A_b$. By this definition, a spherical shape 
has $\Psi = 1$ and distorted shapes have $\Psi < 1$.
We define the bubble as the region 
with $\alpha_g \ge 0.5$ and its surface is the isosurface of $\alpha_g = 0.5$.
We compute $V_b^\prime$ and $A_b$ using high-order interpolation of the data.

\begin{figure}
    \centering
    \includegraphics[width=0.85\columnwidth]{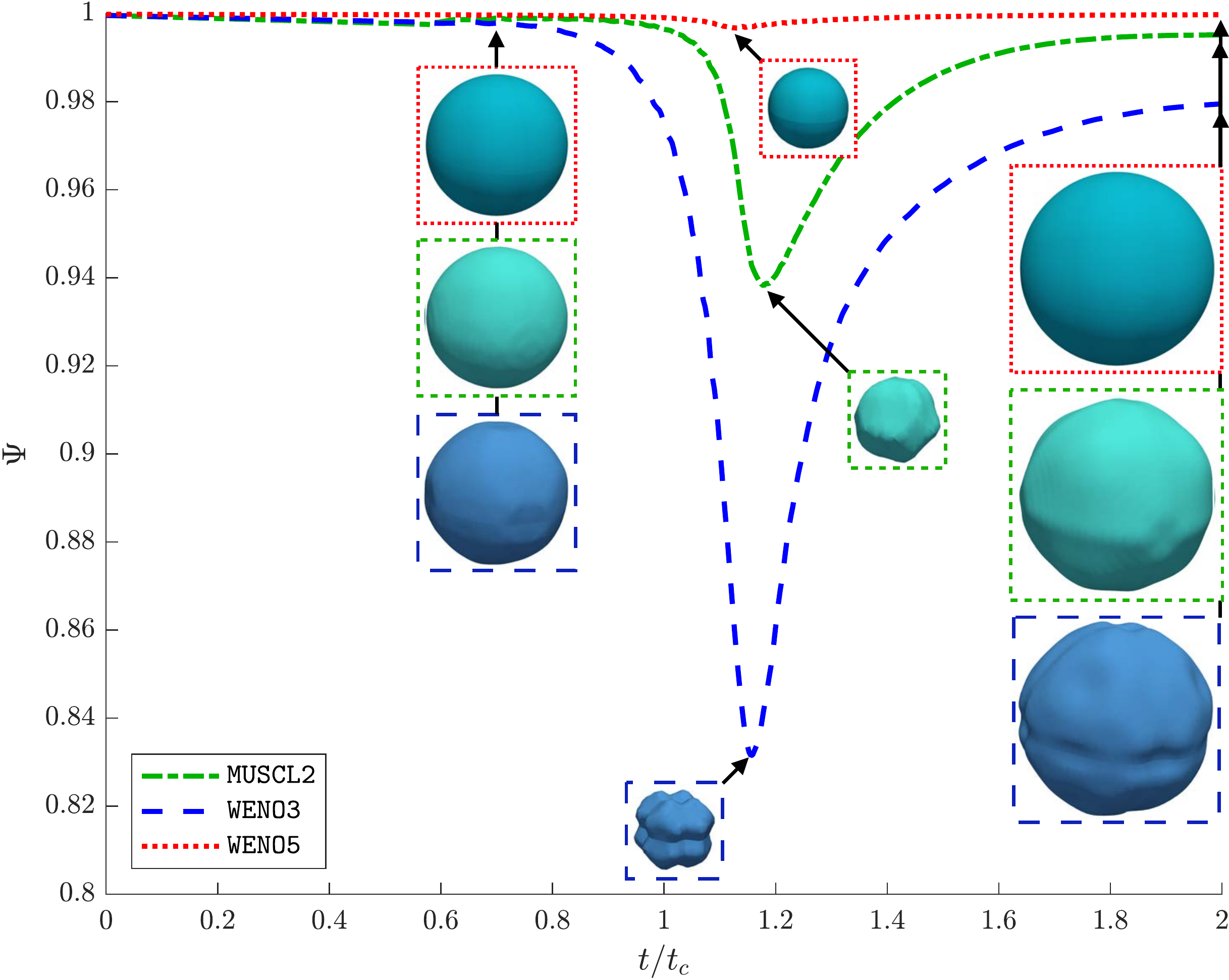}
    \caption{
    Evolution of the bubble sphericity for $p_\infty / p_b = 10$ and $N_{R_0} = 25$.
    Nominal bubble shapes as represented by $\alpha = 0.5$ 
    isosurfaces are also shown for times $t = 0.7 t_c$, 
    $t \left( R = R_{\textrm{min}} \right)$, and $2 t_c$.
    }
    \label{fig_shapesAndSphecity_P10}
\end{figure}

Sphericity and bubble shape evolution for the small pressure ratio
case are shown in Figure~\ref{fig_shapesAndSphecity_P10}. 
We see that the 
\texttt{WENO5} scheme maintains sphericity during the entire 
collapse--rebound process. The \texttt{MUSCL2} and \texttt{WENO3} schemes 
develop grid-specific artifacts, which are visible beginning at $t = 0.7 t_c$;
these are presumably due to anisotropic dispersion on the grid with faster 
propagation of the interface along the Cartesian 
coordinate directions. By the time of minimum radius  $t(R=R_\text{min})$, the 
bubble shape is significantly distorted, and at $t = 2 t_c$
distortions are still visible.

\begin{figure}
    \centering
    \includegraphics[width=0.995\columnwidth]{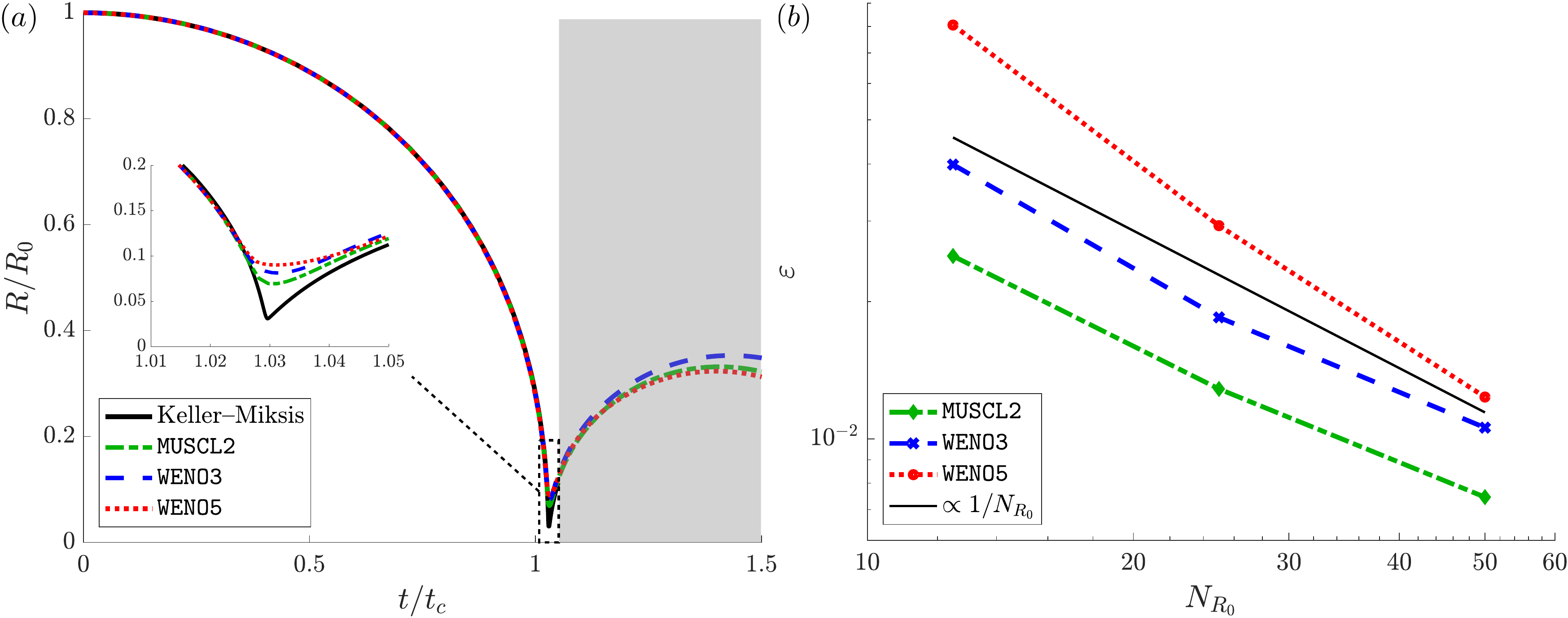}
    \caption{
    Radial bubble-wall evolution for initial interface equilibrium and
    $p_\infty / p_b = 1427$.
    (a) Results (fixed resolution $N_{R_0} = 50$) and
    (b) spatial convergence of the \texttt{MUSCL2}, \texttt{WENO3} and \texttt{WENO5} schemes.
    }
    \label{fig_lowHighOrder_eq}
\end{figure}

The radial bubble-wall evolution and convergence results for the larger pressure ratio $p_\infty / p_b = 1427$ are shown in 
Figure~\ref{fig_lowHighOrder_eq}.
In Figure~\ref{fig_lowHighOrder_eq} (a), we only 
show the Keller--Miksis solution until $t = 1.05 t_c$, just after the 
minimum bubble radius is achieved, 
since the subsequent rebounds for large pressure ratios are well-known
to be physically inaccurate~\citep{fuster2011singleBubble}.
We see that \texttt{MUSCL2} is marginally more accurate at predicting the minimum bubble 
radius and collapse time than the \texttt{WENO5} method. 
This seems to be a result of two factors; first, the interface moves more quickly for larger pressure ratios 
and thus, the \texttt{MUSCL2} results are less polluted by numerical diffusion over the significantly fewer time 
steps to reach collapse than were required for the low pressure-ratio case;
second, the initial smearing introduced for the \texttt{WENO5} method results 
in an initial diffusion greater than that what ultimately develops during 
\texttt{MUSCL2} and \texttt{WENO3} simulations.

In Figure~\ref{fig_lowHighOrder_eq} (b) we plot the observed spatial convergence of the numerical schemes.
Here, we only compute $\eps$ over the temporal window $t \in [0,1.05 t_c]$, commensurate
with the physical accuracy of the Keller--Miksis solution over this interval.
We again observe approximately first-order convergence for all numerical methods we consider. 
However, in this case, \texttt{MUSCL2} has the smallest error $\eps$ and \texttt{WENO5}
the largest. Again, this appears to be a result of the dissipation introduced
by the initial smearing procedure used for the \texttt{WENO5} simulations.


\begin{figure}
    \centering
    \includegraphics[width=0.85\columnwidth]{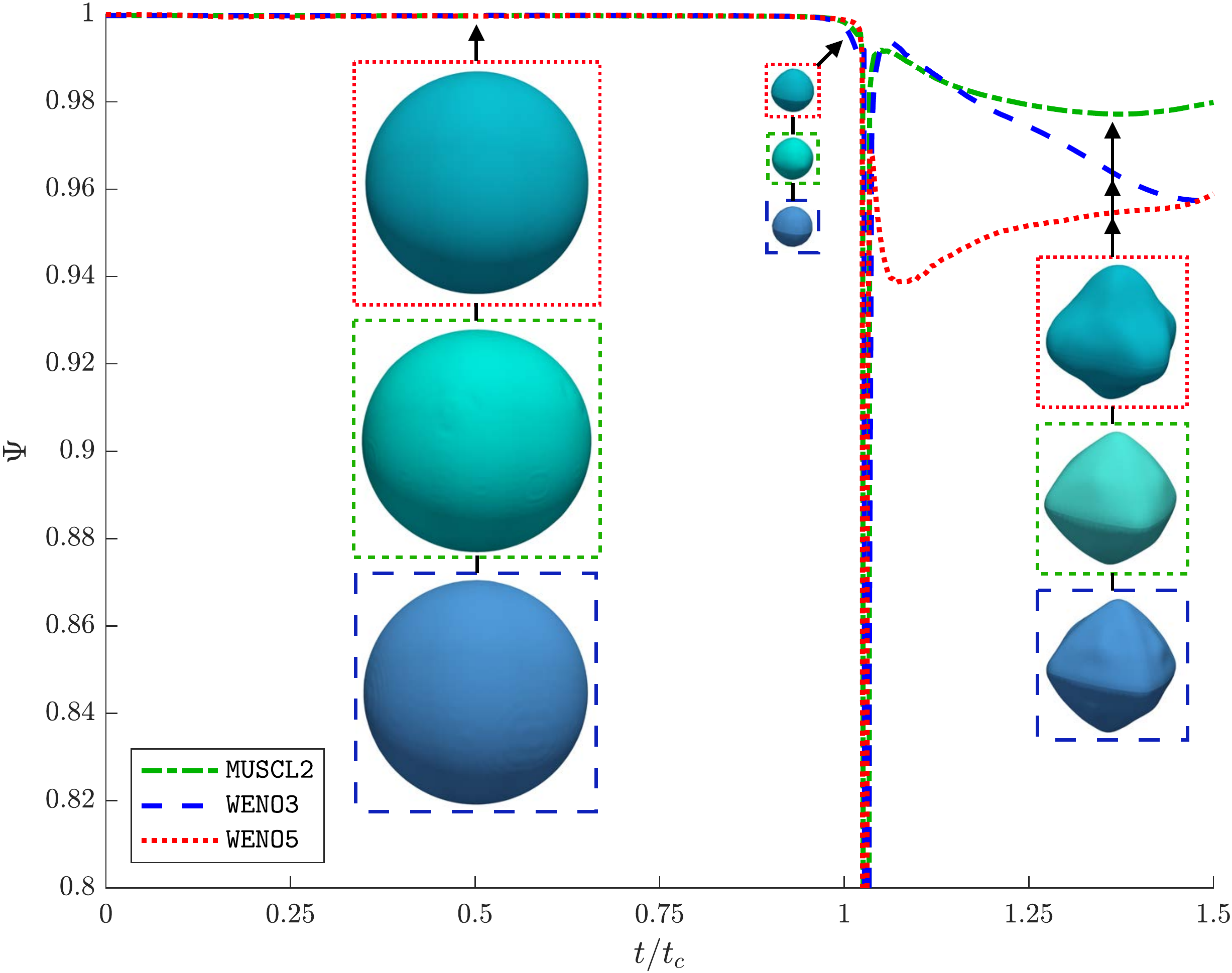}
    \caption{
    Bubble sphericity evolution for $p_\infty / p_b = 1427$ and $N_{R_0} = 50$.
    Nominal bubble shapes ($\alpha = 0.5$)  are also 
    shown for times $t = 0.5 t_c$, 
    $1.02 t_c$, and $1.35 t_c$. 
    }
    \label{fig_shapesAndSphecity_P1427}
\end{figure}

The bubble sphericity and illustrations of the bubble surface 
are shown in Figure~\ref{fig_shapesAndSphecity_P1427}.
Almost no grid-based artifacts on the bubble surface 
are visible until $t \approx t_c$ for all numerical methods,
at which point $\Psi$ decreases significantly. 
Compared to the low-pressure-ratio case, the interface 
evolves more quickly and all methods conserve sphericity for $t \lesssim t_c$. 
However, after the collapse, significant distortions are visible and $\Psi$ does
not reach unity for any of the methods.
Furthermore, we see that the \texttt{WENO5} method results in stronger distortions 
than the \texttt{MUSCL2} or \texttt{WENO3} schemes immediately after the collapse.
For larger $t$, 
the \texttt{WENO3} scheme develops further distortions, eventually
reaching similar $\Psi$ values as \texttt{WENO5} result, whereas 
the \texttt{MUSCL2} scheme maintains sphericity after the initial collapse.

\begin{figure}
    \centering
    \includegraphics[width=0.5\columnwidth]{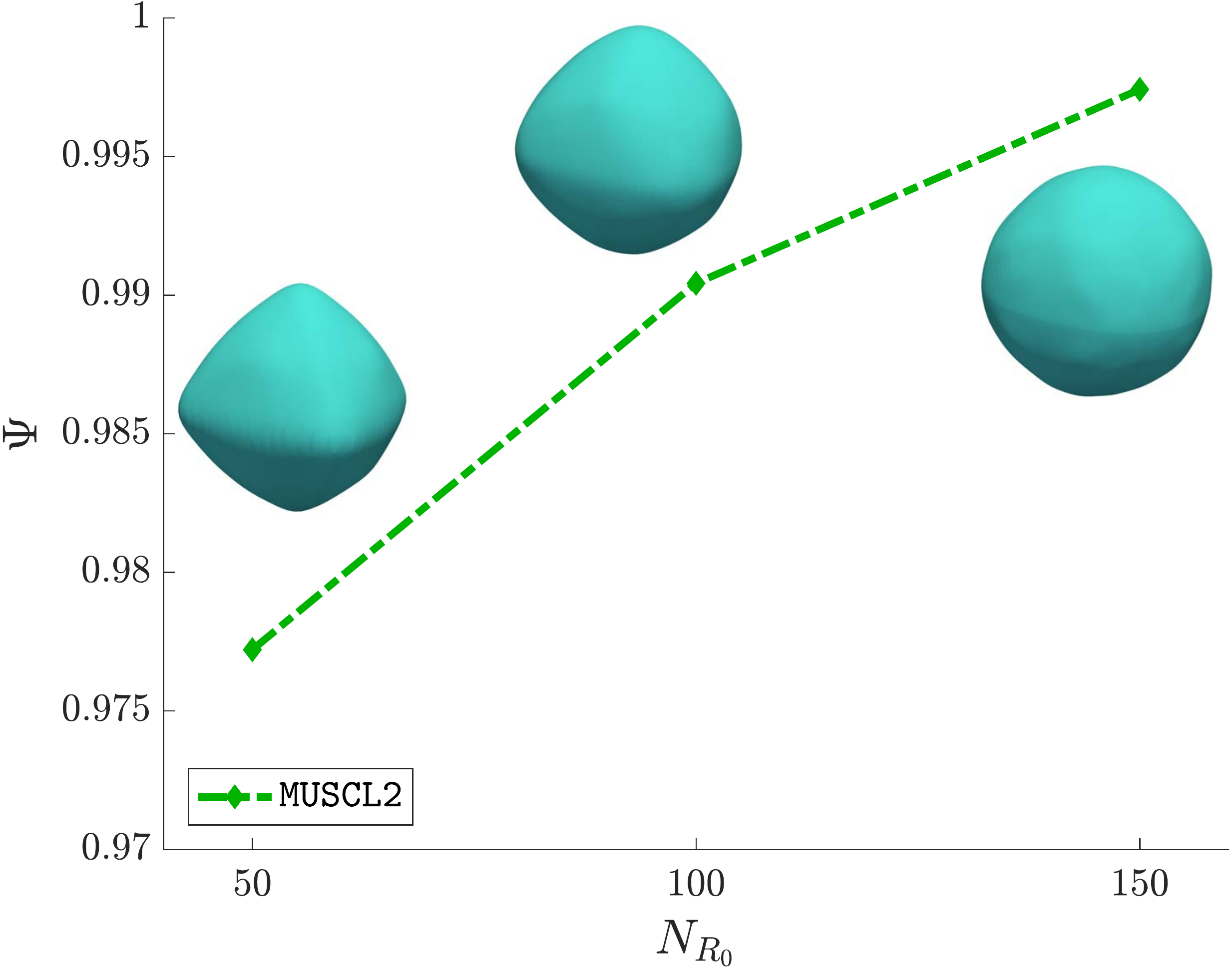}
    \caption{
    Bubble sphericity and associated bubble interface for varying mesh resolution
    $N_{R_0}$ at time $t = 1.35 t_c$ for the $p_\infty / p_b = 1427$
    case and \texttt{MUSCL2} method. For $N_{R_0} = 100$ and 150 the
    adaptive-mesh-refinement technique of~\citet{schmidmayer2018AMR} 
    is used to minimize computational expense.
    }
    \label{fig_shapesAndSphecity_P1427_meshConvergence}
\end{figure}

For $N_{R_0} = 50$, the minimum radius is 
about $0.09R_0$, which corresponds to about 4.5 cells per bubble radius in each direction
and seemingly leads to a significant amount of anisotropy.
Thus, we investigate the effect of mesh resolution on the bubble shape 
in~Figure~\ref{fig_shapesAndSphecity_P1427_meshConvergence}.
We see that sphericity indeed improves with increasing the mesh resolution,
Note that we only shows results for \texttt{MUSCL2}, 
but similar behavior is expected for the WENO schemes.

For initial interface equilibrium, 
we conclude that the \texttt{WENO5} scheme converges more quickly
and can better maintain sphericity when the pressure ratio is relatively
small, and thus the maximum interface velocity is much smaller than the Mach number. 
However, when the pressure ratio is much larger, 
and so the interface velocity exceeds the Mach number,
all the schemes show similar performance, with \texttt{MUSCL2} 
only modestly outperforming the others.

\subsubsection{Spherical bubble collapse with initial interface disequilibrium}\label{sec:results_OutOfEq}

Lastly, we consider the case of initial interface disequilibrium, and thus 
$\dot{R}_0 (t=0) \ne 0$. We enforce this by setting the internal and external 
interface pressures to different values as
\begin{gather}
    p = 
    \begin{cases}
        p_b & \mbox{for } 0 \le R \le R_0 , \\
        p_\infty & \mbox{otherwise.}
    \end{cases}
\end{gather}
This condition represents the discontinuities present,
for example, during bubble wall impact.
The other initial conditions are identical to previous test cases 
and thus of section~\ref{sec:setup}.

\begin{figure}
    \centering
    \includegraphics[width=0.495\columnwidth]{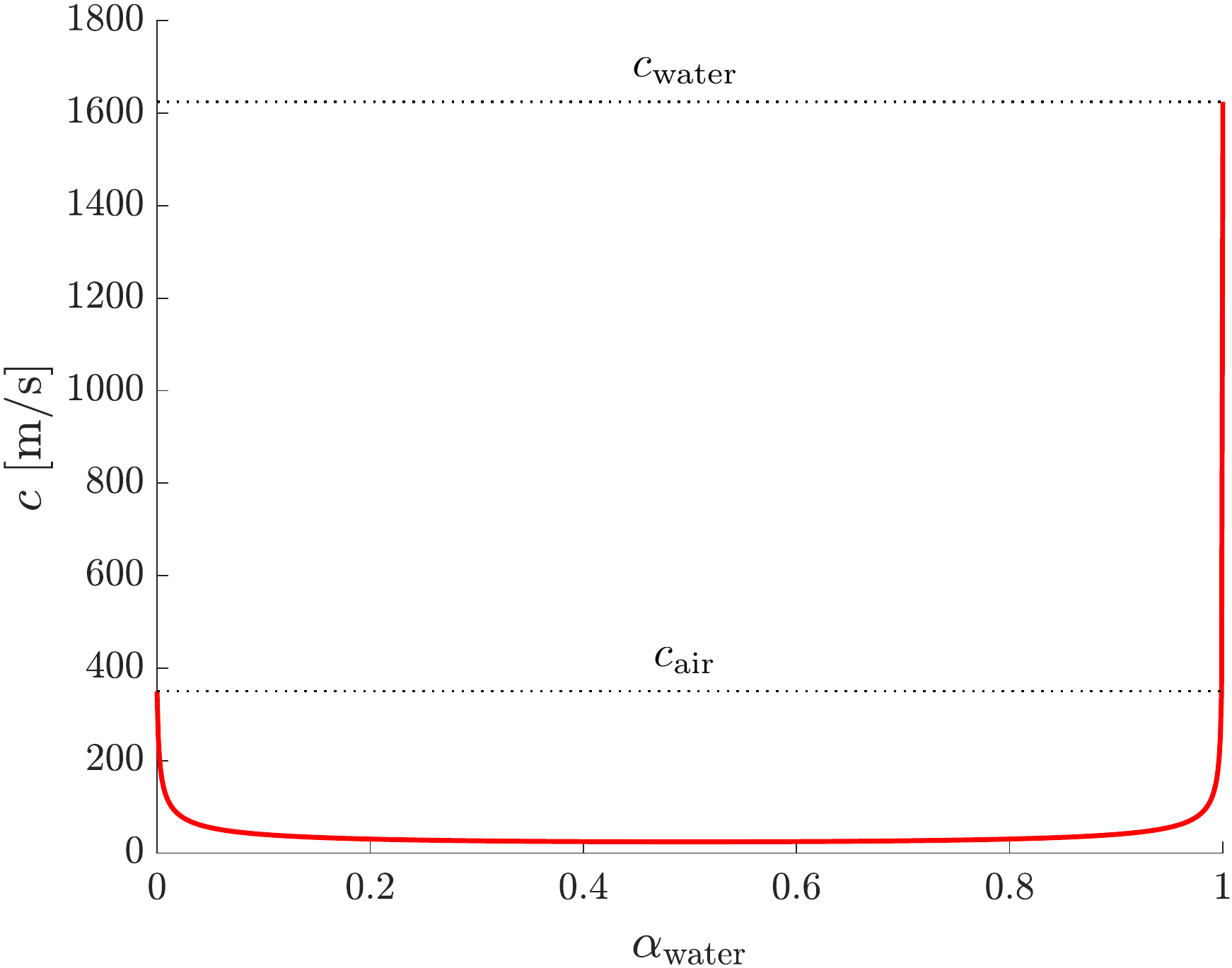}
    \caption{Wood speed of sound for water-air mixture.
    Here, $c_\mathrm{water} = \unit{1625}{\meter\per\second}$ and
    $c_\mathrm{air} = \unit{350}{\meter\per\second}$.
    }
    \label{fig_speedsOfSound}
\end{figure}

With the initial interface smearing employed for the  
\texttt{WENO5} scheme (or indeed after a sufficient number
of time steps for any scheme due to numerical diffusion)
the interface has non-negligible
thickness, giving rise to a mixture region ($\alpha \ne 0$ or 1).
As shown in Figure~\ref{fig_speedsOfSound}, the Wood speed of sound~\eqref{eq_speedOfSound_Wood}, 
which is also the speed of sound of the
5-equation model with $\kdivu$, varies in this region and is much
less than that of either of the pure phases ($\alpha = 0$ and 1). 
After the pressure relaxation
procedure, the effective speed of sound of the 6-equation
model also converges to the Wood speed of sound (see Appendix~\ref{a:vacuum}).

\begin{figure}
    \centering
    \includegraphics[width=0.495\columnwidth]{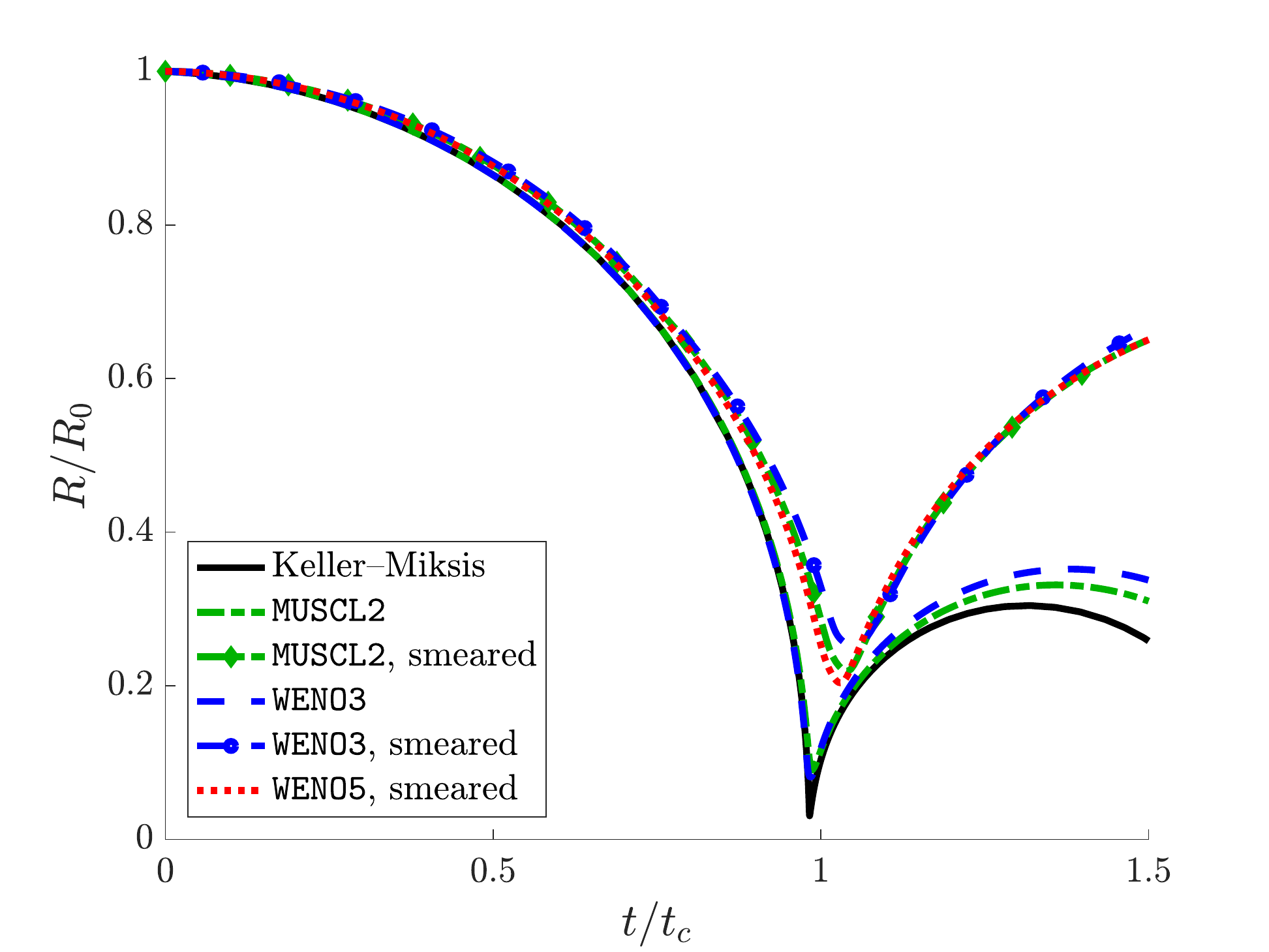}
    \caption{
    Radial bubble-wall evolution for initial interface disequilibrium,
    $p_\infty / p_b = 1427$ and $N_{R_0} = 50$.
    Solutions are computed using the 6-equation model,
    and the Keller--Miksis result is shown as surrogate truth.
    }
    \label{fig_lowHighOrder_outEq}
\end{figure}

Figure~\ref{fig_lowHighOrder_outEq} shows that
the smearing procedure employed to keep the \texttt{WENO5} scheme
stable results in an inaccurate solution for the collapse of a bubble
in initial pressure disequilibrium.  We also see that the \texttt{MUSCL2} and \texttt{WENO3} schemes behave similarly when the
interface is initially smeared, though this procedure is not required
for numerical stability in these cases; for both schemes, the non-smeared cases agree closely with the Keller--Miksis dynamics.

\begin{figure}
    \centering
    \includegraphics[width=0.995\columnwidth]{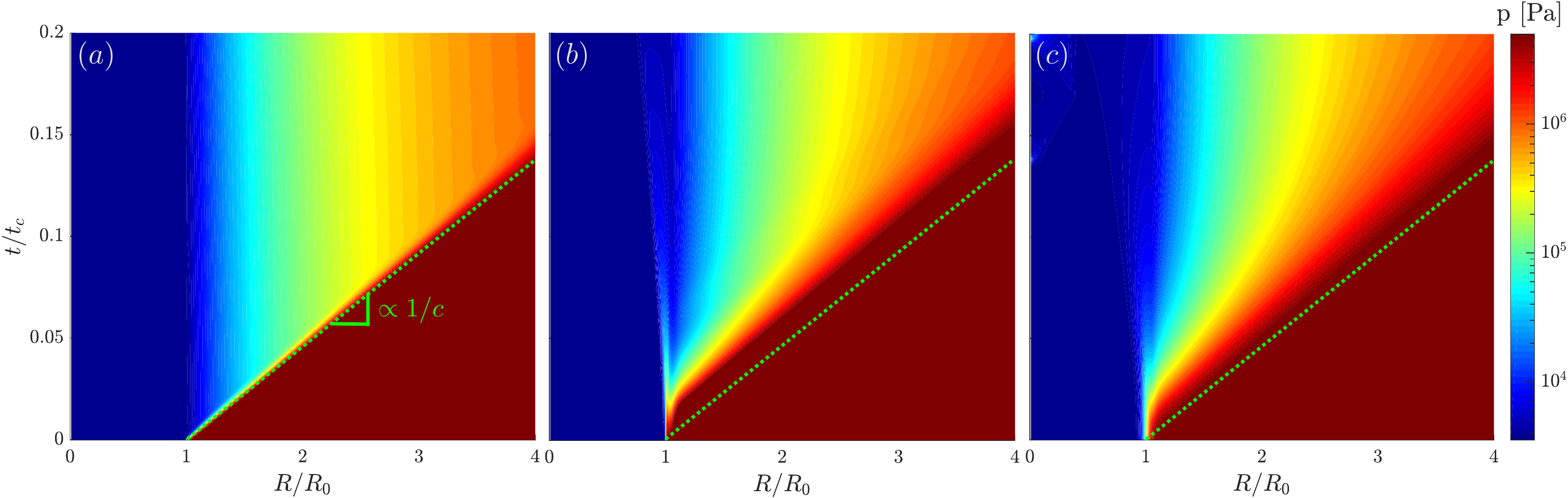}
    \caption{
    Pressure in the time and radial-direction space 
    for initial interface disequilibrium,
    $p_\infty / p_b = 1427$ and $N_{R_0} = 50$.
    Solutions are computed using the 6-equation model
    with the 
    \texttt{MUSCL2} scheme for
    three initial configurations: (a) no initial smearing, 
    (b) smearing on only the volume fraction $\alpha$, and 
    (c) smearing on both $\alpha$ and pressure $p$.
    }
    \label{fig_lowHighOrder_outEq_pressureFields}
\end{figure}

The poor performance associated with the initial interface smearing procedure appears to be due to a 
wave-trapping phenomenon that results from a lower mixture sound speed, 
reducing the initial interface velocity.  This is illustrated in
Figure~\ref{fig_lowHighOrder_outEq_pressureFields} for the \texttt{MUSCL2} method.
Pressure contours are shown in the $t$--$R$ space for three
degrees of initial smearing (a)--(c). 
When the interface is not smeared, the pressure waves travel at 
the pure-phase speed of sound. However, when either the volume fraction
or both the volume fraction and mixture pressure are spatially smeared, these
waves evolve in a more complex manner due to the reduced sound speeds within 
the interface mixture region. Pressure waves that escape the mixture region again travel at
the liquid speed of sound. 
The difference between Figure~\ref{fig_lowHighOrder_outEq_pressureFields}~(b) and 
(c) shows that smearing of the volume fraction $\alpha$ and pressure $p$ 
both modify the pressure-wave behaviors uniquely, though both ultimately
pollute the bubble dynamics.

\subsection{Interface-sharpening techniques for collapsing spherical bubbles}\label{s:THINC}

The numerical dissipation inherent 
in any interface capturing scheme will eventually smear even initially sharp interfaces.
Thus, problems involving multiple interface pressure-disequilibrium events, 
such as a collapsing ellipsoidal bubble near a 
wall~\citep{pishchalnikov2018experimental, pishchalnikov2019kidneyStones}, 
would benefit from keeping interfaces as sharp as possible.

We buttress the 6-equation model and
the \texttt{MUSCL2} and \texttt{WENO3} schemes
with the THINC interface-sharpening method~\citep{shyue2014thinc}
as an illustration of the behavior of interface-sharpening methods
for spherical bubble dynamics.
We use THINC, rather than anti-diffusion~\citep{so2012anti} or
regularization methods~\citep{tiwari2013diffuse}, as its conservative property
matches the conservative methods we already employ and the others 
displayed significant numerical instabilities for our methods
(particularly for high-pressure-ratio cases).
We note that our implementation of the THINC method 
faithfully represented the solutions to the 1D shock tube problem of Appendix~\ref{a:shocktube} and
2D shock--bubble gas--gas interaction problems (such as that
of~\citet{shyue2014thinc} and~\citet{xiao2018THINC2}), and so we can proceed 
with a faithful comparison for collapsing spherical bubbles.
Further, we noticed that the THINC method has numerical instabilities
when coupled to the \texttt{WENO5} scheme; this seems to be a result of
significant interface sharpening, which were previously shown to trigger
instabilities for this method, and so we do not consider \texttt{WENO5} herein.

\begin{figure}
    \centering
    \includegraphics[width=0.995\columnwidth]{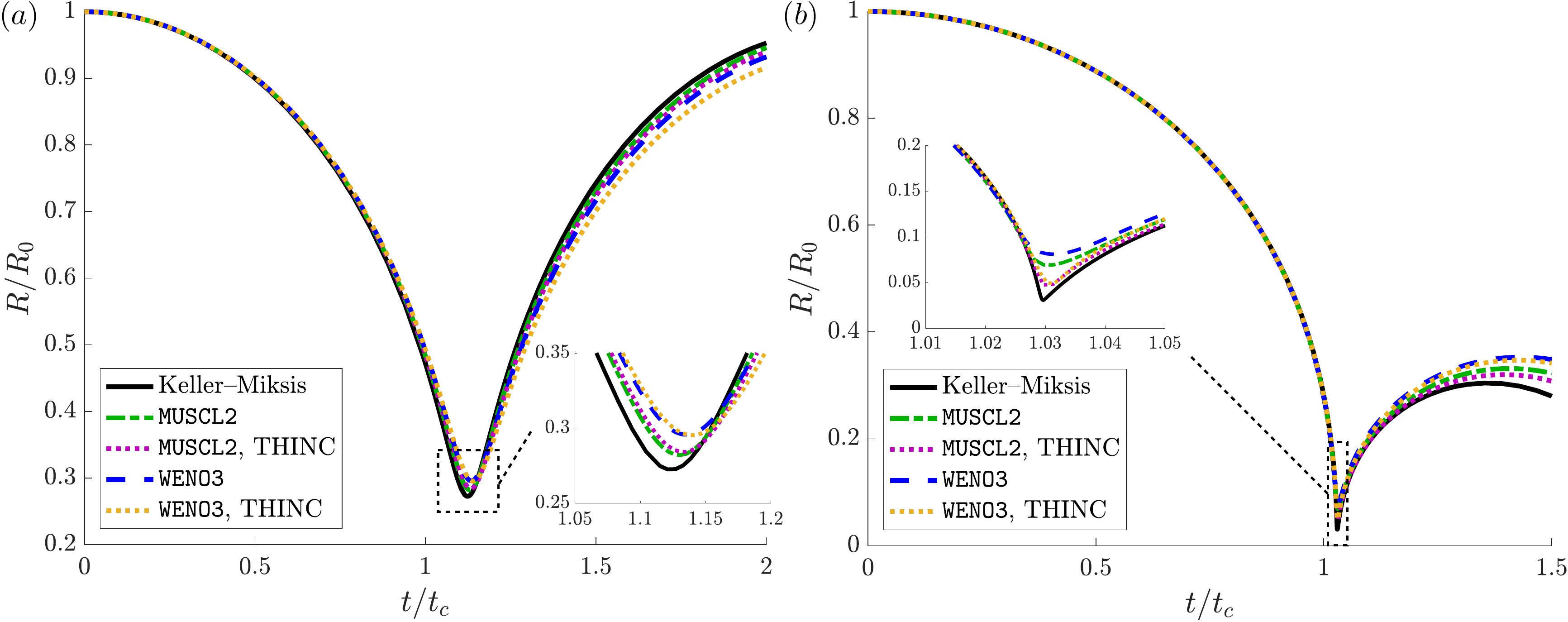}
    \caption{
    Radial bubble-wall evolution for initial interface equilibrium and
    numerical methods as labeled;
    (a) $p_\infty / p_b = 10$ and $N_{R_0} = 25$,
    (b) $p_\infty / p_b = 1427$ and $N_{R_0} = 50$.
    }
    \label{fig_THINC}
\end{figure}

\begin{table}[H]
    \centering
    \begin{tabular}{|l|c|c|c|c|}
        \hline
        \multicolumn{1}{|c|}{Case} & \texttt{MUSCL2} & \texttt{MUSCL2} + THINC & \texttt{WENO3} & \texttt{WENO3} + THINC \\
        \hline
        \textbf{1}: $p_\infty/p_b = 10$ & 0.32 & 0.16 & 0.44 & 0.16 \\
        \hline
        \textbf{2}: $p_\infty/p_b = 1427$ & 0.36 & 0.1 & 0.5 & 0.08 \\
        \hline
    \end{tabular}
    \caption{Interface thickness $T/R_0$ at specific times and for the cases as labeled.
    Case \textbf{1}: $N_{R_0} = 25$ and $t = 2 t_c$;
    case \textbf{2}: $N_{R_0} = 50$ and $t = 1.35 t_c$
    Here, $T$ is computed via the number of cells satisfying
    $0.01 \leqslant \alpha \leqslant 0.99$.}
    \label{t:interface_sharpness}
\end{table}

Figure~\ref{fig_THINC} shows the radial bubble-wall evolution
for the low- and high-pressure-ratio interface-pressure 
equilibrium cases we considered in 
section~\ref{sec:results_methods_interfaceEquilibrium}.
When compared to non-THINC-equipped methods, 
the THINC results have about $44\%$ larger error $\eps$ for 
the low-pressure-ratio case and $59\%$ smaller $\eps$ for the 
high-pressure-ratio case; however, in all cases the error is already relatively small.
Despite having an inconsistent effect on the error, the THINC scheme does
keep the bubble interface sharper, as shown in Table~\ref{t:interface_sharpness}.

\begin{figure}
    \centering
    \includegraphics[width=0.995\columnwidth]{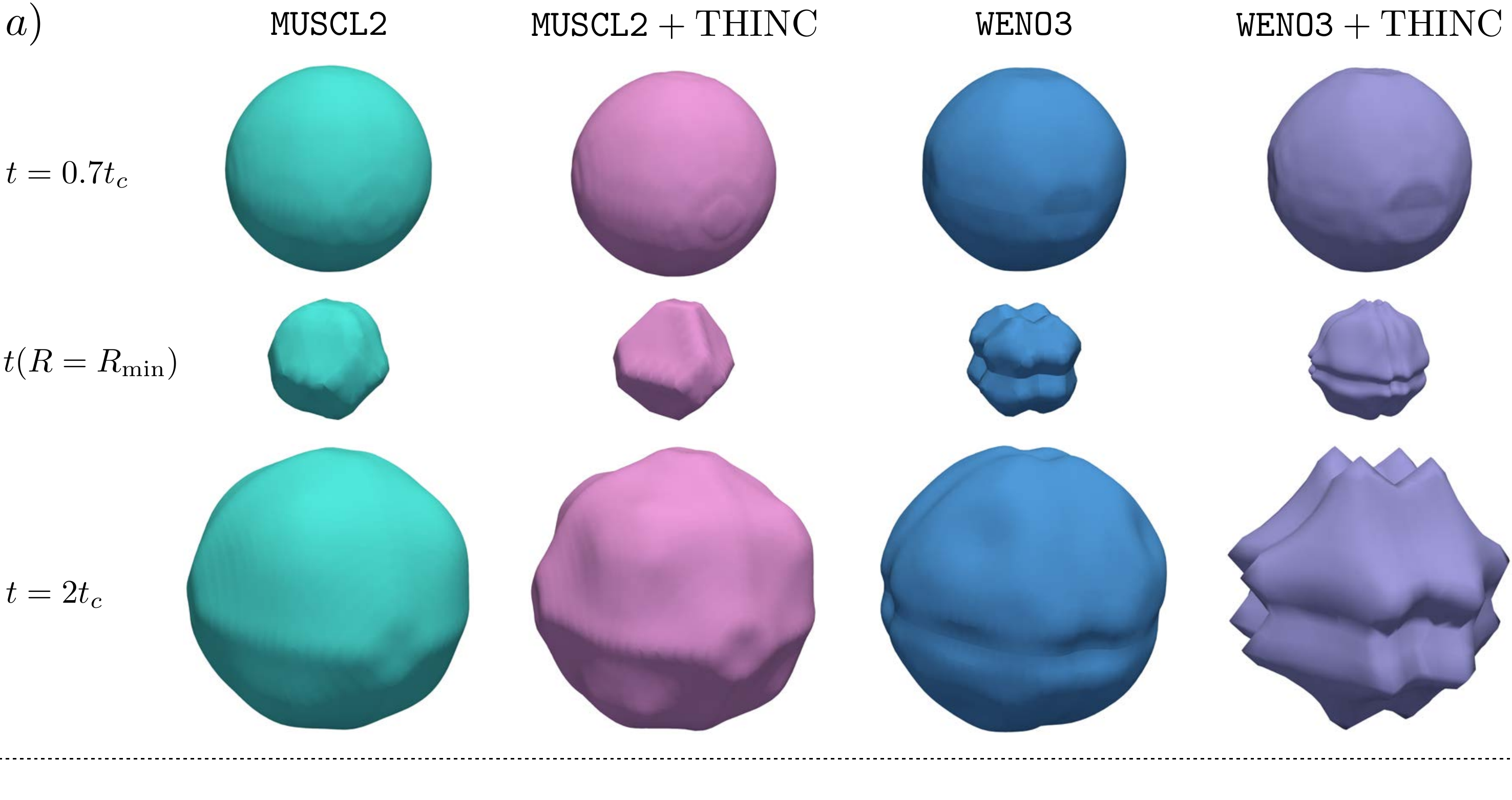}
    \includegraphics[width=0.995\columnwidth]{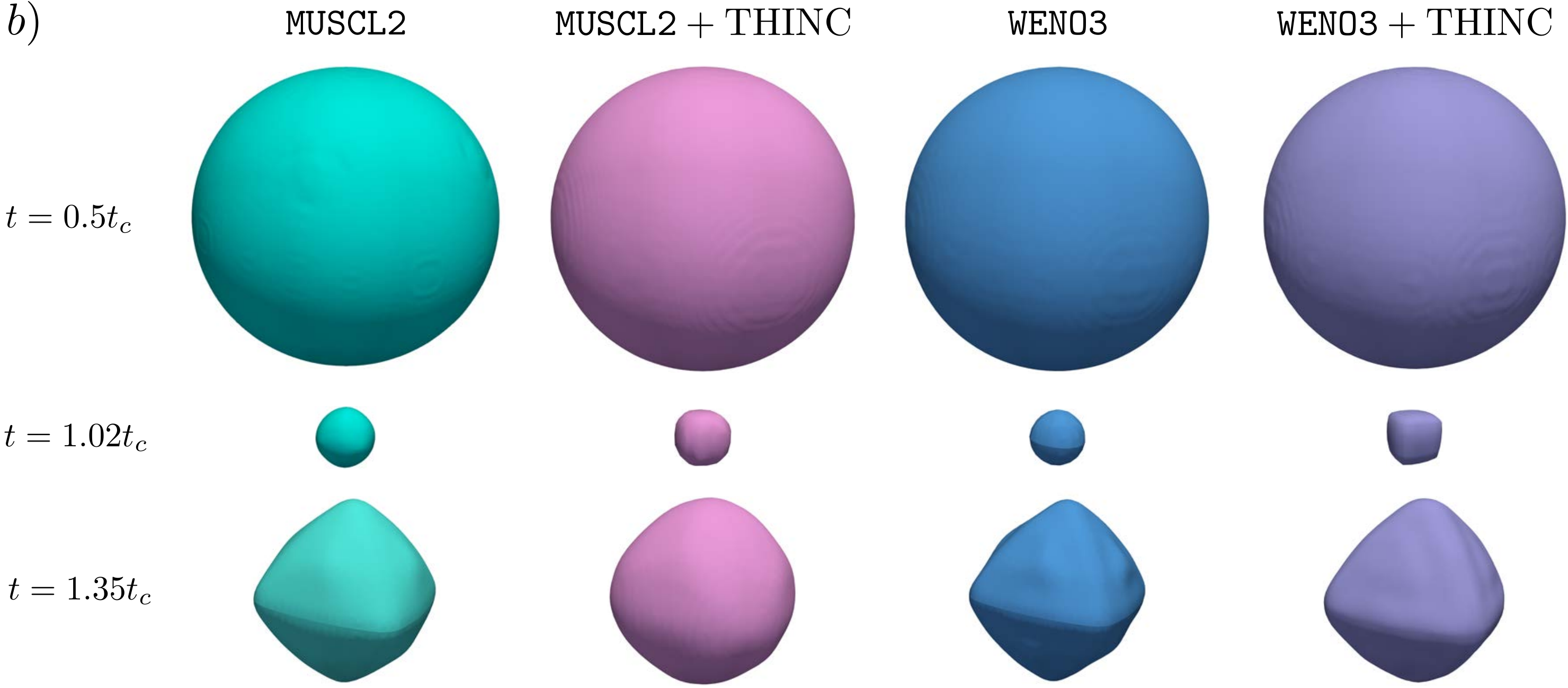}
    \caption{
    Nominal bubble shapes ($\alpha = 0.5$) for times and methods as labeled:
    (a) $p_\infty / p_b = 10$ and $N_{R_0} = 25$,
    (b) $p_\infty / p_b = 1427$ and $N_{R_0} = 50$.
    }
    \label{fig_THINC_shapes}
\end{figure}

Figure~\ref{fig_THINC_shapes} compares bubble shapes with and without THINC. 
We see that the THINC method results in significantly less 
spherical shapes for the low-pressure-ratio cases,
though for the high-pressure-ratio cases the shapes are nearly the same.
In general, we see that the THINC method is better behaved when coupled
to the \texttt{MUSCL2}, rather than the \texttt{WENO3}, scheme.

Thus, we conclude that the THINC method did not reliably improve the accuracy of our
results, and in most cases disturbed the interface sphericity. As such, it only offers a 
partial solution when considering collapsing bubbles 
with multiple pressure-disequilibrium events.

\section{Discussion and conclusion}\label{sec:Conclusion}

We analyzed the ability of diffuse-interface models and their associated 
numerical methods to represent the collapse and
rebound of spherical gas bubbles in a liquid. We confirmed that the 5-equation model 
of~\citet{allaire} is unable to accurately represent a spherical bubble collapse 
and demonstrated how the additional $\kdivu$ term introduced by~\citet{kapila2001}
is required to ensure good agreement with the Keller--Miksis solution \citep{tiwari2015growth}.
Since the 5-equation model with $\kdivu$ is known to produce instabilities in some numerical
experiments~\citep{relaxjcp, beig2018temperatures}, we investigated the 6-equation pressure-disequilibrium model as a potential
surrogate. We observed good agreement between these models for challenging
test problems, including a 1D water-air shock tube, 
a 1D vacuum developing in a water-air mixture, and the collapse of a 3D spherical bubble. 
Thus, the 6-equation model is a good candidate to remedy the stability issues of the
5-equation model with the $\kdivu$ source term.

We also considered the behavior and pathologies of the 6-equation model 
when coupled to MUSCL and WENO numerical methods for a collapsing
spherical bubble. We first analyzed bubbles at initial interface pressure 
equilibrium. For this, the bubble interface evolution of the \texttt{WENO5}-based solution
more closely matched the associated Keller--Miksis surrogate-truth solution than did the
\texttt{MUSCL2} and \texttt{WENO3} schemes for relatively small pressure ratios.
This was due to the more substantial numerical diffusion intrinsic to 
the lower-order schemes, and despite the fact that the \texttt{WENO5} scheme
required an initially smeared interface to maintain simulation stability.
When the initial pressure ratio was larger, all three methods showed similar 
results, quickly converging to the Keller--Miksis solution. 
Further, we noticed that the relatively small bubble size at the collapse
time resulted in significantly distorted interface shapes. 
However, these shapes were shown to be more spherical for finer spatial meshes.
Thus, an adaptive-mesh-refinement technique would be helpful for maintaining bubble
interface sphericity at the same computational cost as a uniform mesh near the bubble.

When the bubble interface was in initial disequilibrium,
we saw that the smearing procedure implemented for the \texttt{WENO5}
method precluded an accurate solution for large pressure ratios. 
This was a result of the relatively large degree of initial diffusion,
which produced a mixture region with a much smaller speed of sound
that polluted the dynamics. We also noted that the numerical dissipation inherent 
in any interface capturing scheme will eventually smear even initially sharp interfaces
and, therefore, these schemes 
would benefit from keeping interfaces as sharp as possible.
Interface-sharpening techniques are one way to minimize this dissipation,
and we surveyed the THINC method~\citep{shyue2014thinc} for the
same spherical bubble collapse problems.
While the THINC method did keep the interfaces sharper, in most cases it
further disturbed the interface sphericity; additionally, we did not observe a 
consistent increase in simulation accuracy. Thus, further
investigation and possibly method improvement are required to maintain surface
sharpness while guaranteeing a conservative behavior and numerical stability.

Ultimately, we saw that WENO-based schemes were preferable for bubble 
dynamics that involve small pressure ratios, and thus slower interface dynamics,
and the MUSCL and WENO-based schemes performed similarly for 
large pressure ratios and thus fast interface speeds. Thus, 
the \texttt{WENO5} scheme is generally preferred, except in cases
involving interface pressure discontinuities, for which the interface
smearing required to keep the scheme stable pollutes the dynamics.
As such, the instability of high-order WENO schemes for interface problems
warrants future attention.




\section*{Acknowledgments}

The authors would like to thank Dr.\ Mauro Rodriguez, Prof.\ Eric Johnsen, and 
Dr.\ Shahaboddin Alahyari Beig for fruitful discussions. This work was supported by the Office of 
Naval Research under grant numbers N0014-18-1-2625 and N0014-17-1-2676.

\bibliography{Biblio}{}
\bibliographystyle{elsarticle-num-names}

\begin{appendices}
\newpage

\section{Water-air shock tube}\label{a:shocktube}

\begin{figure}
    \centering
    \includegraphics{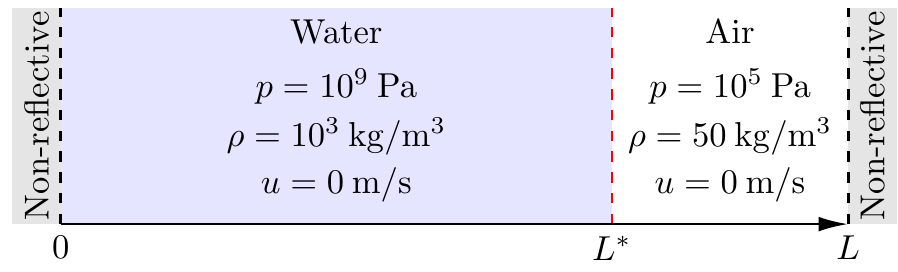}
    \caption{Shock tube problem setup and initial and boundary conditions.}
    \label{f:shocktube_setup}
\end{figure}

We consider a water-air shock tube, whose initial configuration
is shown in Figure~\ref{f:shocktube_setup}~\citep{relaxjcp, schmidmayer2018AMR, beig2015maintaining}. 
The domain has length $L$, the initial discontinuity 
is located at $L^* = L/7$ and $10^3$ nodes are used.
Here, the water has 
stiffened-gas parameters $\gamma_l = 4.4$ and 
$\pi_\infty = \unit{6 \times 10^8}{\pascal}$~\citep{murrone2005fiveEq, relaxjcp, schmidmayer2017capillary}.

\begin{figure}
    \centering
    \includegraphics[width=0.96\columnwidth]{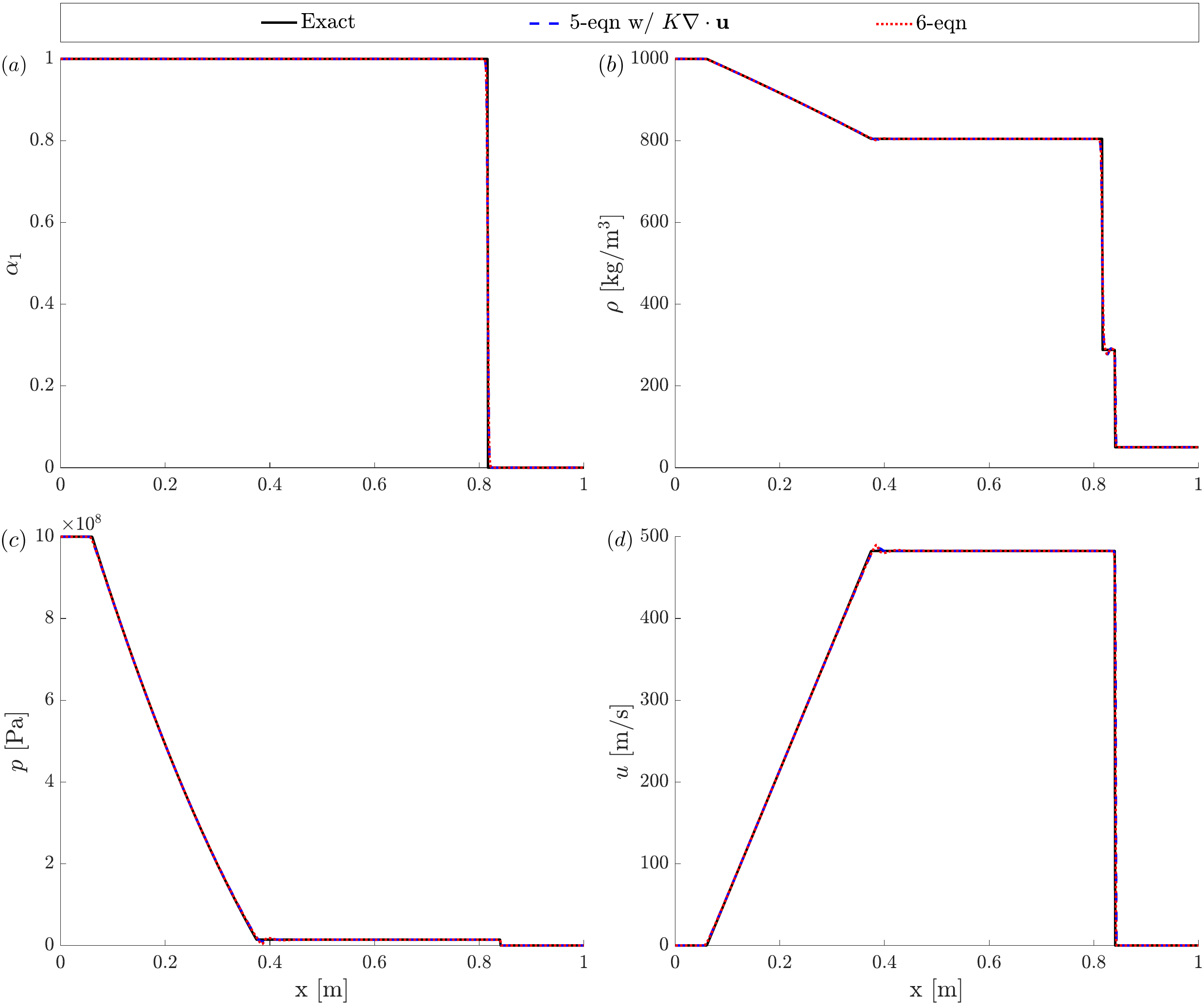}
    \caption{
    Water-air shock-tube interaction problem at $t = \unit{241}{\micro\second}$. Numerical and exact solutions are as labeled above.
    }
    \label{f:shocktube}
\end{figure}

We simulate the flow in the shock tube using both the 5-equation with $\kdivu$
and 6-equation models and the \texttt{WENO5} numerical scheme. 
A uniform and one-dimensional mesh of $10^3$ nodes is used.
Results for the primitive variables at $t = \unit{241}{\micro\second}$ are shown in Figure~\ref{f:shocktube}.
A rightward shock wave propagates into the air,
followed by a contact discontinuity, observable
in (a) and (b), that delimits the interface
between the two phases; left-going
expansion waves propagate into the water.
We observe good agreement between the numerical implementations of both models and the exact solution.
Indeed, differences can only be seen at the tail of the expansion waves 
and near the contact discontinuity. Note that these differences diminish with increasing resolution.

\section{Vacuum generation into a water-air mixture}\label{a:vacuum}

\begin{figure}
    \centering
    \includegraphics{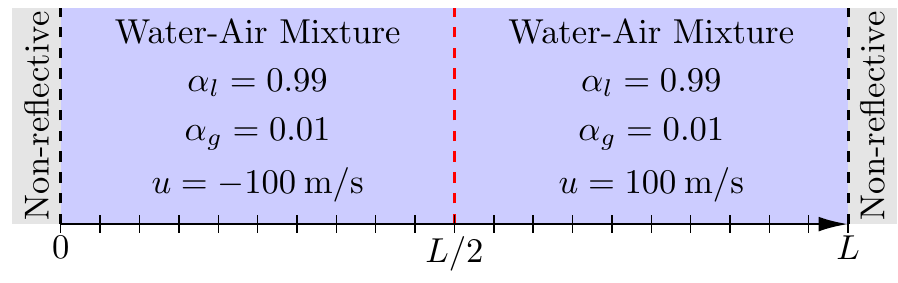}
    \caption{Problem setup for a vacuum generation into a water-air mixture.}
    \label{f:vacuum_setup}
\end{figure}

We consider a vacuum generation into a water-air mixture~\citep{relaxjcp, pelanti2014mixture}.
The problem setup is shown in Figure~\ref{f:vacuum_setup};
there is a uniform initial pressure $p = \unit{10^5}{\pascal}$
and densities $\rho_{l} = \unit{10^3}{\kilo\gram\per\meter\cubed}$ 
and $\rho_{g} = \unit{1}{\kilo\gram\per\meter\cubed}$,
and a flow is generated by the initial discontinuity in velocity.
Again, $10^3$ nodes are used and the water has stiffened-gas parameters $\gamma_l = 4.4$ and 
$\pi_\infty = \unit{6 \times 10^8}{\pascal}$.

\begin{figure}
    \centering
    \includegraphics[width=0.96\columnwidth]{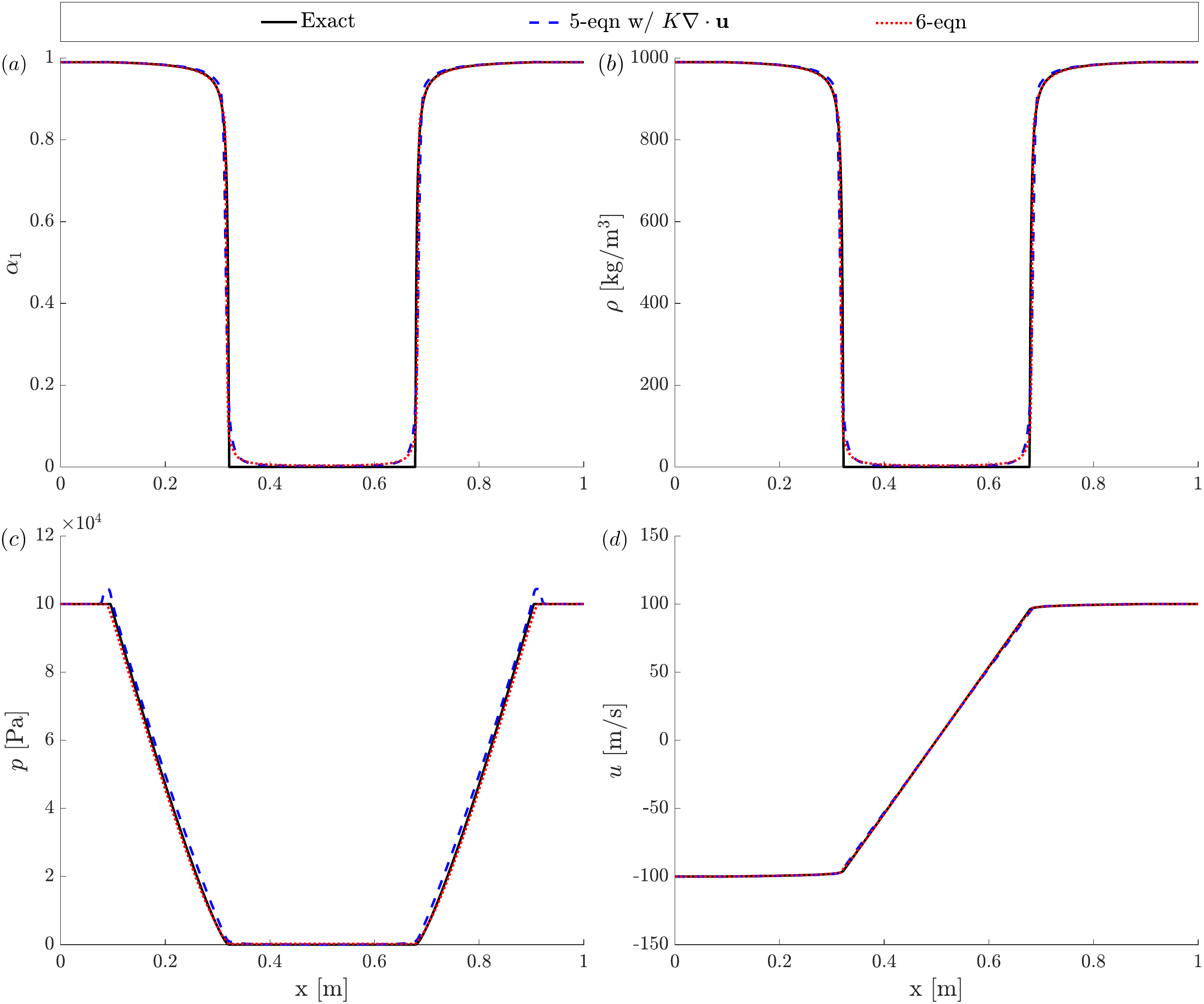}
    \caption{
    Cavitating water-air mixture problem at $t = \unit{1.85}{\milli\second}$. Numerical and exact solutions are as labeled above.
    }
    \label{fig_vacuum}
\end{figure}

Figure~\ref{fig_vacuum} shows the results of the primitive variables 
at $t = \unit{1.85}{\milli\second}$ for 
the vacuum problem using the same methods 
and computational parameterization as Appendix~\ref{a:shocktube}.
The discontinuity in velocity generates left- and right-going expansion waves, and thus
generates a $p=0$ vacuum in the center of the domain.
Mixture compressibility ensures that the water volume fraction, and thus the mixture density,
decreases in the vacuum region. We observe good agreement 
between the numerical simulations and exact solution.
However, the 6-equation model generally performs better, 
with no pressure oscillations at the head of the expansion waves.

\end{appendices}

\end{document}